\documentclass[aps,pra,reprint,amsmath,amssymb,showpacs,groupedaddress,floatfix]{revtex4-1}
\usepackage{mathtools,graphicx,bbold,bm,color}
\usepackage[colorlinks,citecolor=blue,urlcolor=blue,linkcolor=blue,hyperindex,breaklinks]{hyperref}
\usepackage{ulem,multirow}

\definecolor{astral}{RGB}{46,116,181}
\definecolor{azulclaro}{RGB}{32,119,199}
\definecolor{cinves}{RGB}{18,144,129}
\definecolor{guinda}{RGB}{128,23,14}

\begin{document}

\title{Instabilities in an optical black-hole laser}
\author{Juan David Rincon-Estrada and David Bermudez}
\email[E-mail: ]{dbermudez@fis.cinvestav.mx}
\homepage[\\ Website: ]{https://www.fis.cinvestav.mx/~dbermudez/}
\affiliation{Departamento de F\'isica, Cinvestav, A.P. 14-740, 07000\ Ciudad de M\'exico, Mexico}

\date{\today}

\begin{abstract}
The Hamiltonian of optical fields in a nonlinear dispersive fiber is studied. Quantum field fluctuations are spontaneously created close to an optical event horizon through the analog Hawking effect. The simplest model is considered for an optical black-hole laser, where the Hawking radiation is produced and amplified inside a cavity formed by two horizons: a black hole and a white hole. It is found that resonant Hawking radiation originates from a discrete set of instabilities and tunnels out of the horizons. Finally, the numerical results are compared with the resonance and instability conditions and a phenomenological model is developed to give a clear physical picture.

PhySH: quantum aspects of black holes, laboratory studies of gravity, optical fibers, Kerr effect.
\end{abstract}

\maketitle

\section{Introduction}\label{intro}
Hawking radiation is the flux of particles with thermal spectrum that escapes from a black hole \cite{Hawking1974}. It is considered a fundamental phenomenon of quantum field theory in curved spacetimes and used as a test for possible theories of quantum gravity \cite{Helfer2003,Visser2003}. Although its formulation is based on widely accepted physical principles, its astrophysical observation seems unlikely, at least in the near future.

In 1981, Unruh realized that the same effect occurs in moving fluids \cite{Unruh1981}. This opened the possibility of observing analog Hawking radiation in a laboratory instead of directly from an astrophysical black hole. In a moving fluid, the analog phenomenon occurs close to a sonic horizon---the interface that separates a subsonic and a supersonic current. A thermal flux of phonons is spontaneously generated close to the sonic horizon and into the subsonic region. This effect ultimately originates from the impossibility of defining a global vacuum state that is adapted to both regions of the fluid.

The seminal work of Unruh inspired further proposals to test predictions of gravitation and cosmology in laboratories. The most common analog systems in hydrodynamics, condensed matter, and optics are: water tanks \cite{Weinfurtner2011,Euve2016}, liquid helium \cite{Volovik2003,jacobson1998event}, Bose-Einstein condensates (BECs) \cite{zapata2011resonant,2018Bermudez,de2019observation}, optical fibers \cite{philbin2008fiber,drori2019observation}, and spintronic materials \cite{jannes2011hawking}. A new configuration emerging from these studies is the so-called ``black-hole laser'' \cite{Corley1999}, that consists of a stationary finite transonic fluid: its flow changes from subsonic to supersonic, and after a finite region back to subsonic. This configuration establishes two horizons---one corresponding to a black hole (BH) and the other one to a white hole (WH)---that confine Hawking radiation in the region between them. The radiation is self-amplified because the two horizons behave as mirrors in a resonant cavity---similar to a laser. This process exists in sonic analogs (as BECs) if the quantum field is bosonic and the dispersion relation is anomalous. The same effect was extended to the optical case to obtain an ``optical black-hole laser'' (OBHL) for a bosonic field and normal dispersion relation \cite{Faccio2012,GaonaReyes2017}.

The theory of instabilities describes this amplification process \cite{Leonhardt2003, Hydrodynamic} by analyzing the quantum excitations spontaneously generated in a moving medium. This method was previously applied in condensed matter \cite{Macher2009pra,2018Bermudez,Finazzi2010njp,Coutant2010} and in this work we implement it in optics, where the laboratory and comoving frequencies take the roles of the wavenumber and the frequency, respectively. We use the flat velocity profile approximation and obtain information that is hidden in the technical difficulties of more complicated models commonly used in optics: Hawking radiation escapes from the cavity, the resonant modes originate from a discrete set of instabilities in the system, and the cavity modes have different lifetimes.
This work is organized as follows. In Section \ref{sec2model}, we start from the Hamiltonian for light in an optical fiber and review the derivation of the dynamical equation for a quantum field and the dispersion relation for a spontaneous fluctuation. In Section \ref{sec3obhl}, we analyze the OBHL configuration with real frequencies and describe its kinetics. We obtain the instabilities of the system by introducing complex frequencies in Section \ref{sec4insta}. In Section \ref{sec5compa}, we compare these results with the resonance and instability conditions and introduce a phenomenological model to expand the physical picture. Finally, we present our conclusions in Section \ref{sec6conclu}.

\section{Quantum fluctuations propagating in an optical fiber}\label{sec2model}
When an electromagnetic wave travels inside a dielectric medium, it interacts with the bound electrons in the material, such that its group velocity is less than the speed of light and depends on the optical frequency of the wave $\omega$. This dependence is encoded in the material dispersion. If the medium is an optical fiber, we also consider the nonlinear coupling between the electromagnetic wave and the electron polarization. Our goal is to analyze the dynamics of a quantum field fluctuation propagating inside an optical fiber and generated close to the optical horizon established by a light pulse through the analog Hawking effect.

Let us start from the Hamiltonian for a quantum field of light in an optical fiber proposed by Drummond \cite{drummond1990,drummond2014quantum} and given by $H=H_\text{L}+H_\text{NL}$, where $H_\text{L}$ is the linear part of the Hamiltonian
\begin{align}
H_\text{L}=&\hbar \int  \left[\omega\Psi^{\dagger}\Psi
+\frac{i}{2}u\left(\partial_x\Psi^{\dagger}\Psi-\Psi^{\dagger}\partial_x\Psi\right)\right. \nonumber\\
&\left. -\frac{1}{2}u^3\beta_2\partial_x\Psi^{\dagger}\partial_x\Psi \right] \text{d}x,
\end{align}
and $H_\text{NL}$ is the nonlinear part
\begin{equation}
H_\text{NL}=\chi_e\int \left(\Psi^{\dagger}\right)^2\Psi^2 \text{d}x,
\end{equation}
$u=\beta_1^{-1}=(\partial_{\omega}\beta)^{-1}|_{\omega_\text{c}}$ is the group velocity of the field, $\beta_2=\partial^2_\omega \beta|_{\omega_\text{c}}$ is the group-velocity dispersion (GVD) parameter, $\beta$ is the wavenumber, $\omega_\text{c}$ is the central frequency, and $\chi_e$ characterizes the nonlinear coupling of the field with itself.

The quantum field $\Psi(x,t)$ propagates inside an optical fiber and satisfies the following commutation relation
\begin{equation}\label{ec:conmutador}
[\Psi(x,t),\Psi^{\dagger}(y,t)]=\delta(x-y).
\end{equation}
Considering the Hamiltonian in the interaction picture, this quantum field obeys the following equation of motion
\begin{equation}\label{ecscdesp}that
(\partial_t+u\partial_x)\Psi=\left[-\frac{i}{2}u^3\beta_2\partial_x^2+i\chi_e\Psi^{\dagger}\Psi\right]\Psi.
\end{equation} 
The first term in Eq. \eqref{ecscdesp} suggests the following coordinate transformation
\begin{equation}\label{trans1}
\tau=t-\frac{x}{u},\qquad x'=x,
\end{equation}
that introduces a reference frame comoving with the pulse. The change of coordinates yields
\begin{equation}\label{ecsustituir}
\partial_x=\partial_{x'}- \frac{1}{u}\partial_{\tau}, \qquad \partial_{t}=\partial_{\tau}.
\end{equation}
Substituting Eqs. \eqref{ecsustituir} in Eq. \eqref{ecscdesp}, and considering the conditions $|\partial_{x'}^2\Psi|\ll |\partial_{\tau}^2\Psi|$ and $|\partial_{\tau}\partial_{x'}\Psi|\ll |\partial_{\tau}^2\Psi|$, which are commonly fulfilled in optics\cite{drummond2014quantum}, we find
\begin{equation}\label{ec:NLST}
i\partial_{x'}\Psi-\frac{\beta_2}{2}\partial_{\tau}^2\Psi+\chi_e\beta_1\Psi^{\dagger}\Psi\Psi=0.
\end{equation}
This is a nonlinear Schr\"odinger equation (NLSE) where the time and space variables have been interchanged. This equation describes the propagation of a quantum field in a nonlinear medium ($\Psi^{\dagger}\Psi\Psi$) with dispersive effects $(\partial_{\tau}^2\Psi)$.

\subsection{Linearization of the quantum fluctuation}
Consider a pump field propagating inside an optical fiber described by Eq. \eqref{ec:NLST}. If this field is a continuous wave, it does not depend on $\tau$ as it is stationary during propagation. Therefore, we can propose
\begin{equation}\label{ec:fondo1}
\Psi_0(x')=\sqrt{n}\text{e}^{i\kappa x'},
\end{equation}
where $n=|\Psi_0|^2$ is the photon density and $\kappa=\chi_e \beta_1 n$. Let us see if this solution is stable under a small fluctuation $\phi(x',\tau)$, that is, $\Psi\rightarrow\Psi'(x',\tau)=\Psi_0(x')+\phi(x',\tau)\text{e}^{i\kappa x'}$ such that
\begin{equation}
\Psi'(x',\tau)=\text{e}^{i\kappa x'}\left[\sqrt{n}+\phi(x',\tau)\right].\label{antz}
\end{equation} 
Inserting Eq. \eqref{antz} in Eq. \eqref{ec:NLST}, we find the equation for the propagation of a linear fluctuation $\phi$ under a background classical field $\Psi_0$
\begin{equation}\label{ec:difcampo}
i\partial_{x'}\phi-\frac{\beta_2}{2}\partial_{\tau}^2\phi+\kappa\left(\phi+\phi^{\dagger}\right)=0.
\end{equation}
The quantum fluctuation $\phi$ fulfills the following commutation relation
\begin{equation}
\left[\phi(x',\tau),\phi^{\dagger}(y',\tau)\right]=\delta(x'-y').
\end{equation}
As is usually done in quantum field theory, we expand the fluctuation $\phi$ in modes in the wavenumber $\beta$ as
\begin{equation}\label{expbeta}
\phi(x',\tau)=\int d\beta\left(\phi_{\beta}(x',\tau) a_{\beta}+\eta_{\beta}^*(x',\beta) a_{\beta}^{\dagger}\right),
\end{equation}
where the modes satisfy the normalization relations
\begin{align}
\int dx'\left(\phi_{\beta}^*\phi_{\beta'}-\eta^*_{\beta}\eta_{\beta'}\right)&=\delta(\beta-\beta'),\\
\int dx'  \left(\phi_{\beta}\phi_{\beta'}-\eta_{\beta}\eta_{\beta'}\right)&=0,
\end{align}
and $a_{\beta}[a_{\beta}^{\dagger}]$ is the annihilation [creation] operator. This is a consequence of the fundamental commutator in Eq. \eqref{ec:conmutador}.

Using the mode expansion of Eq. \eqref{expbeta}, Eq. \eqref{ec:difcampo} can be written in matrix form as
\begin{equation}
\begin{pmatrix}	
i\partial_{x'}-\frac{\beta_2}{2}\partial_{\tau}^2+\kappa & \ \ \kappa\\ 
\ \ \kappa & -i\partial_{x'}-\frac{\beta_2}{2}\partial_{\tau}^2+ \kappa
\end{pmatrix}\begin{pmatrix}
\phi_{\beta}\\
\eta_{\beta}
\end{pmatrix}=\begin{pmatrix}
0\\
0
\end{pmatrix}.
\end{equation}
The off-diagonal terms mix positive and negative frequency (or norm) modes---an essential ingredient of the Hawking process \cite{Robertson2012jpb}. From this, we obtain the differential equation for $\phi_\beta$
\begin{equation}
\left(\partial_{x'}^2-\beta_2\kappa\partial_{\tau}^2+\frac{\beta_2^2}{4}\partial_{\tau}^4\right)\phi_{\beta}=0.
\label{waveeq}
\end{equation}
Assuming a plane-wave solution $\phi_{\beta}\propto \text{e}^{i(\beta x'-\omega\tau)}$, we find the dispersion relation
\begin{equation}\label{dispori}
\beta^2=\beta_2 \kappa \omega^2+\frac{\beta_2^2}{4}\omega^4=\frac{\omega^2}{c^2}\left(b_1^2+b_2\omega^2\right),
\end{equation}
where $b_1^2=c^2\beta_2 \kappa$ and $b_2=c^2\beta_2^2/4$ are given by the material constants and the group velocity of the pulse. This is a normal dispersion, because the wavenumber $\beta(\omega)$ increases with $\omega$, as is common in optics \cite{GaonaReyes2017}.

It is useful to replace the propagation distance $x'$ for a propagation time $\zeta=x'/u$, such that the frame comoving with the pulse is described by the coordinates $(\zeta,\tau)$. The conjugated variable to $\zeta$ is the comoving frequency $\omega'$ and it is a conserved quantity \cite{Bermudez2016pra,Robertson2012jpb}. The comoving frequency can be obtained from the Doppler shift as
\begin{equation}\label{ecdoppler}
\omega'=\gamma[\omega-u\beta(\omega)]=\gamma\omega\left(1-\frac{n(\omega)}{n_{g0}}\right),
\end{equation}
where $\gamma=(1-u^2/c^2)^{-1/2}$, $n_{g0}=c/u$, and we used the relation $\beta(\omega)=n(\omega)\omega/c$ to replace the wavenumber $\beta(\omega)$ for the refractive index $n(\omega)$. Together with the dispersion relation \eqref{dispori}, we obtain 
\begin{equation}\label{nnew}
n(\omega)=\sqrt{b_1^2+b_2\omega^2}.
\end{equation}

Considering an optical pulse as background, $n(\omega)$ has an additional contribution $\delta n$ due to the optical Kerr effect. We define the effective refractive index as $n_\text{eff}(\omega)=n(\omega)+\delta n$. By replacing $n(\omega)\rightarrow n_\text{eff}(\omega)$ in Eq. \eqref{ecdoppler}, using Eq. \eqref{nnew}, and rearranging terms, we find a new dispersion relation $\omega'(\omega)$ in the comoving frame
\begin{equation}\label{ecdisoptica}
\left[\frac{n_{g0}}{\gamma b_1}\omega'+\frac{v(\tau)}{b_1}\omega\right]^2=\omega^2+\frac{\omega^4}{\Omega_0^2},
\end{equation}
where $\Omega^2_0=b_1^2/b_2$, and we define the velocity profile of the optical medium as
\begin{equation}\label{ec:perfilvel}
v(\tau)=-n_{g0}+\delta n(\tau),
\end{equation}
that is, a change of the group index. We refer to this dimensionless quantity as a velocity in analogy with the dispersion relation in fluids \cite{Corley1999}. According to Eq. \eqref{ec:perfilvel}, $v(\tau)$ is always negative and the medium flows to the left. The configuration is stationary in the comoving frame and the change in velocity is due to the change in refractive index $\delta n$.
To simplify the notation, without loss of generality we set $b_1=1$ and absorb the constant $n_{g0}/\gamma$ in $\omega'$. Then, the dispersion relation for the optical case in Eq. \eqref{ecdisoptica} becomes
\begin{equation}\label{disperoptica}
\left[\omega'+v(\tau)\omega\right]^2=\omega^2+\frac{\omega^4}{\Omega_0^2}.
\end{equation}
There are two branches of this dispersion relation
\begin{equation}
\omega'=\omega\left(-v(\tau)\pm\sqrt{1+\frac{\omega^2}{\Omega_0^2}}\right),
\end{equation}
where the $+$ [$-$] sign corresponds to the counterpropagating or u-modes [copropagating or v-modes]. As is customary, the labels of counterpropagating and copropagating come from their motion in the laboratory frame with respect to the effective flow. We will see that the conditions for the Hawking effect lead to the change of direction of travel in the comoving frame for one mode.

The general solution for a mode of the quantum fluctuation with fixed $\omega'$ is
\begin{equation}\label{econdaplana2}
\phi(\zeta,\tau)=A\text{e}^{-i[\omega'(\omega)\zeta+\omega \tau]}.
\end{equation}
The dispersion relation \eqref{disperoptica} can also be obtained directly from the following differential equation
\begin{equation}
\left(\left(\partial_{\zeta}+v\partial_{\tau}\right)^2-\partial_{\tau}^2+\frac{1}{\Omega_0^2}\partial_{\tau}^4\right)\phi=0,
\end{equation}
that is obtained directly from an action in Ref. \cite{GaonaReyes2017}. We can obtain this equation by performing the change to coordinates $\tau$ and $\zeta$ directly in Eq. \eqref{ecscdesp}. However, this derivation is less clear given the nontrivial proposal of the pump field in place of Eq. \eqref{ec:fondo1}: The global phase is proportional to the dimensionless velocity profile $v(\tau)$ in Eq. \eqref{ec:perfilvel}.
\subsection{Optical analog of the event horizon}
In fluid analogs of the event horizon, the moving medium represents the black-hole spacetime and waves in the medium represent light waves. In optics, the analogy goes one step further: waves are indeed light waves, but a light pulse propagating in a dielectric replaces the moving medium as the black-hole spacetime \cite{Bermudez2016pra}.

To create an optical horizon, we send a laser pulse with optical power $I(\tau)$---called pump---that slows down waves. This is a consequence of the increase in the local refractive index of the medium due to the Kerr effect by
\begin{equation}\label{n2}
\delta n(\tau)=n_2 I(\tau).
\end{equation}
The resulting refractive index profile establishes the effective spacetime curvature where quantum fluctuations propagate. If light is slowed down below the pump speed, the pump moves superluminally and two horizons are formed at the boundaries between sub- and superluminal propagation: light cannot enter the trailing edge of the pulse or cannot escape from the leading edge of the pulse. In analogy with spacetime metrics, the leading edge of the pulse acts as a black hole (BH) horizon and the trailing edge as a white hole (WH) horizon \cite{jacquet2019analytical, philbin2008fiber}. We show the horizons created by a soliton (sech${^2}$) pulse in Fig. \ref{fig1diagram}(a), this is the most common configuration in optical experiments \cite{philbin2008fiber,drori2019observation}.

The Hawking effect is defined as the spontaneous creation of particles around a black-hole horizon due to the mixing of positive and negative frequency modes. In the optical analog, the field is the electromagnetic field and the horizon is formed by light traveling in a dielectric. The study of analog systems has given us more freedom to consider the Hawking effect in different configurations that cannot be obtained in astrophysics, such as the black-hole laser.

\section{Optical black-hole laser}\label{sec3obhl}
\begin{figure}
	\centering
	\includegraphics[width=0.49\textwidth]{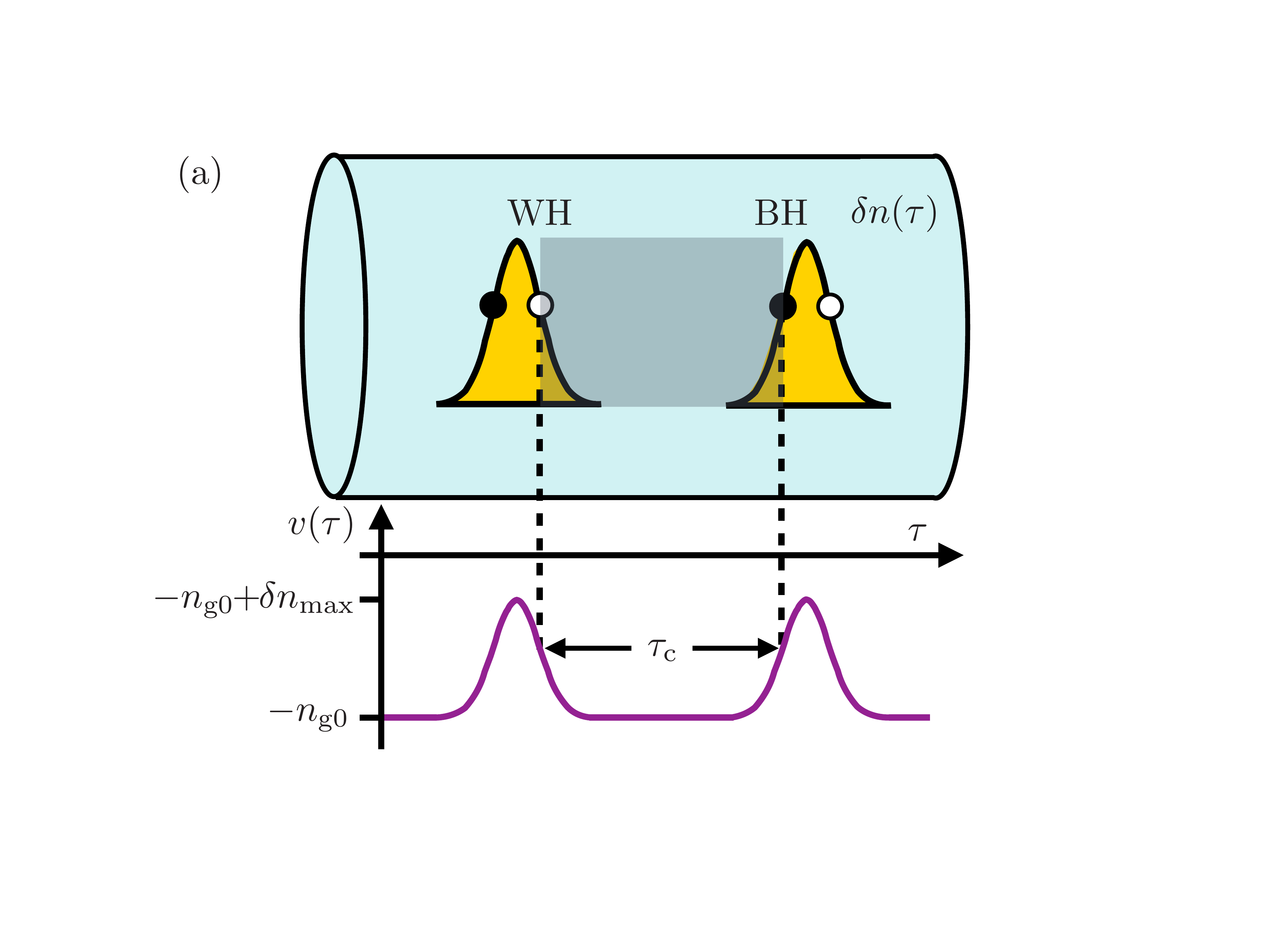}\hspace{1mm}
	\includegraphics[width=0.49\textwidth]{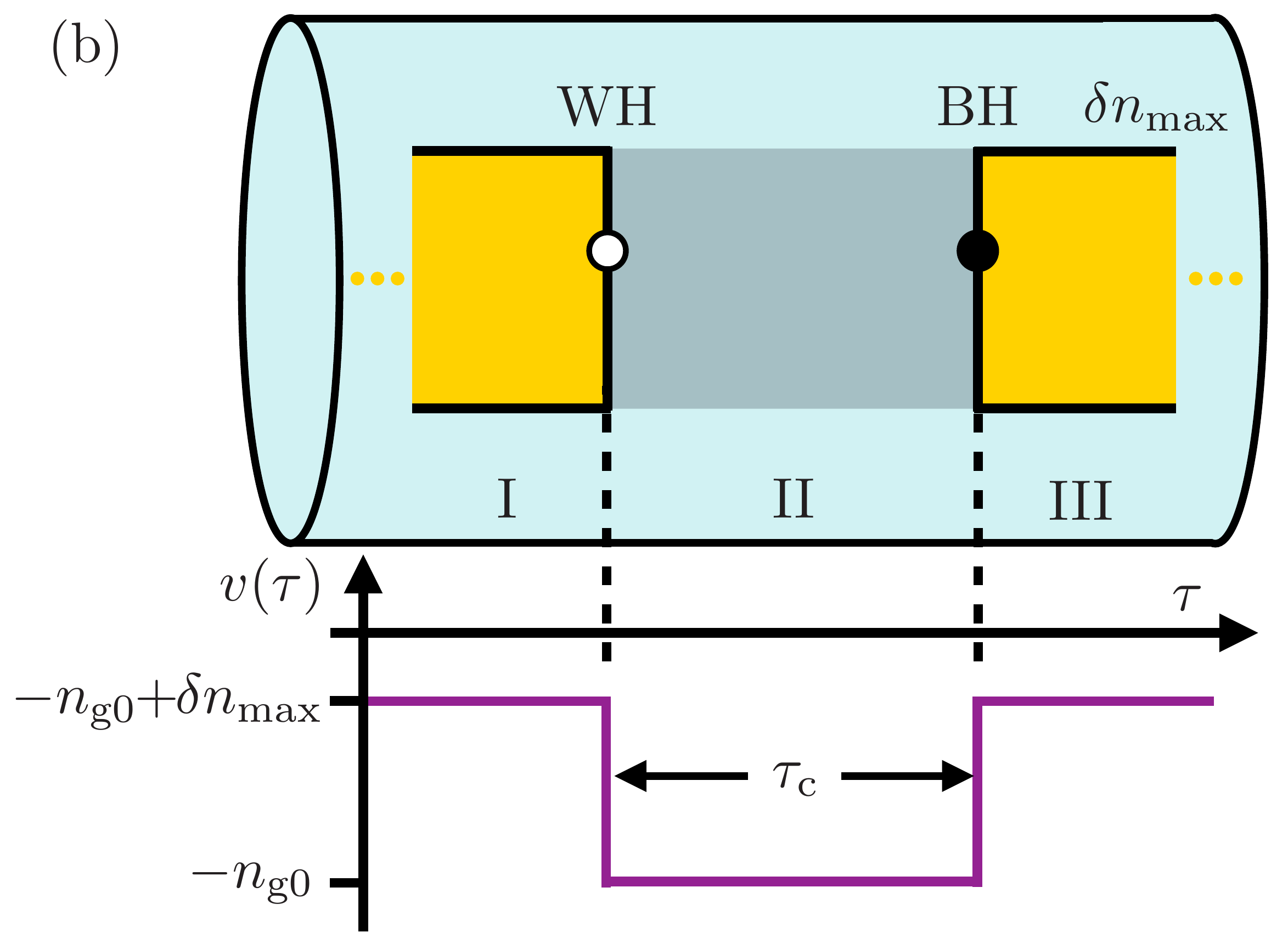}
	\caption{Cavities formed in the comoving frame by two solitons (a) and in the step-index model (b), where the white- (WH) and black- (BH) hole horizons are marked. We show the corresponding velocity profile $v(\tau)$ below each cavity, marking three regions: I and III are the subluminal regions or exterior, and II is the superluminal region or cavity.} 
	\label{fig1diagram}
\end{figure}

In an optical black-hole laser (OBHL), the analog Hawking effect generates a quantum fluctuation that is then trapped and amplified in a cavity made by light. Two light pulses separated by a delay time $\tau_\text{c}$ create an optical cavity in the region between them. If both pulses have the same frequency $\omega_\text{c}$, their locations in the comoving frame remain constant. Each pulse has a BH and a WH, but only the inner horizon of each pulse forms the cavity \cite{GaonaReyes2017,philbin2008fiber}, as shown in Fig. \ref{fig1diagram}(a): The cavity is the shaded region between the WH in the leading pulse and the BH in the trailing one.

The concept of the OBHL inspired the so-called temporal waveguide, where light is trapped, not in space as in a waveguide, but in time between two pulses \cite{Plansinis2016}. The temporal waveguide has been proposed as a way of increasing data speed transmission through optical fibers \cite{Demircan2011}.

In a dispersive system, the group velocity depends on the frequency and, because of this, each frequency experiences the horizon in a different spatial point: the horizon is ``fuzzy''. Dispersion complicates the theoretical analysis of the OBHL. For this reason, we use a simplified model where the cavity is formed, not by solitons, but by sharp changes of refractive index. This is the so-called step-index model \cite{Jacquet2015} or front-induced \cite{jacquet2019analytical}, where the cavity is a region with $\delta n=0$ and the exterior with $\delta n=\delta n_\text{max}$ (see Fig. \ref{fig1diagram}(b)).

The two main differences between the soliton and step-index models are how fast and for how long is the change in the refractive index. In the step-index model, the refractive index is constant and semi-infinite in each side, its change is discontinuous. In the soliton model, the change is continuous and is related to pulse duration. The steepness of a pulse is the analog surface gravity, so ultra-short (and steep) pulses are used to produce analog Hawking radiation \cite{drori2019observation}. However, the containment of the cavity in the OBHL improves for longer (and less steep) pulses. A supergaussian pulse can be used to obtain both behaviors. The approximations in the step-index model increase the production of analog Hawking radiation by increasing the analog surface gravity and the containment of the cavity.

Although both models fulfill the definition of an OBHL, the step-index is simpler and is the one we use for the theoretical analysis in this work. The horizons still create analog Hawking radiation inside and outside the cavity and the cavity amplifies the trapped modes.

\subsection{Numerical method: OBHL with real frequencies}
The step-index model defines two different regions for $v(\tau)$ in Eq. \eqref{ec:perfilvel}: a subluminal region outside the cavity (I and III), where $v_1=-n_{g0}+\delta n_\text{max}$, and a superluminal region that forms the cavity (II), where $v_2=-n_{g0}$. Notice that $v(\tau)$ is negative because in the comoving frame the medium moves to the left. In the optical case, we can only vary the velocity in the outside region $v_1$, while $v_2$ is kept fixed, unlike the sonic case, where the flow in any region can be modified \cite{Leonhardt2007}. This is because the velocity profile in the optical case is modified by $\delta n$, which is always positive.

\begin{figure}
	\centering
	\includegraphics[width=0.49\textwidth]{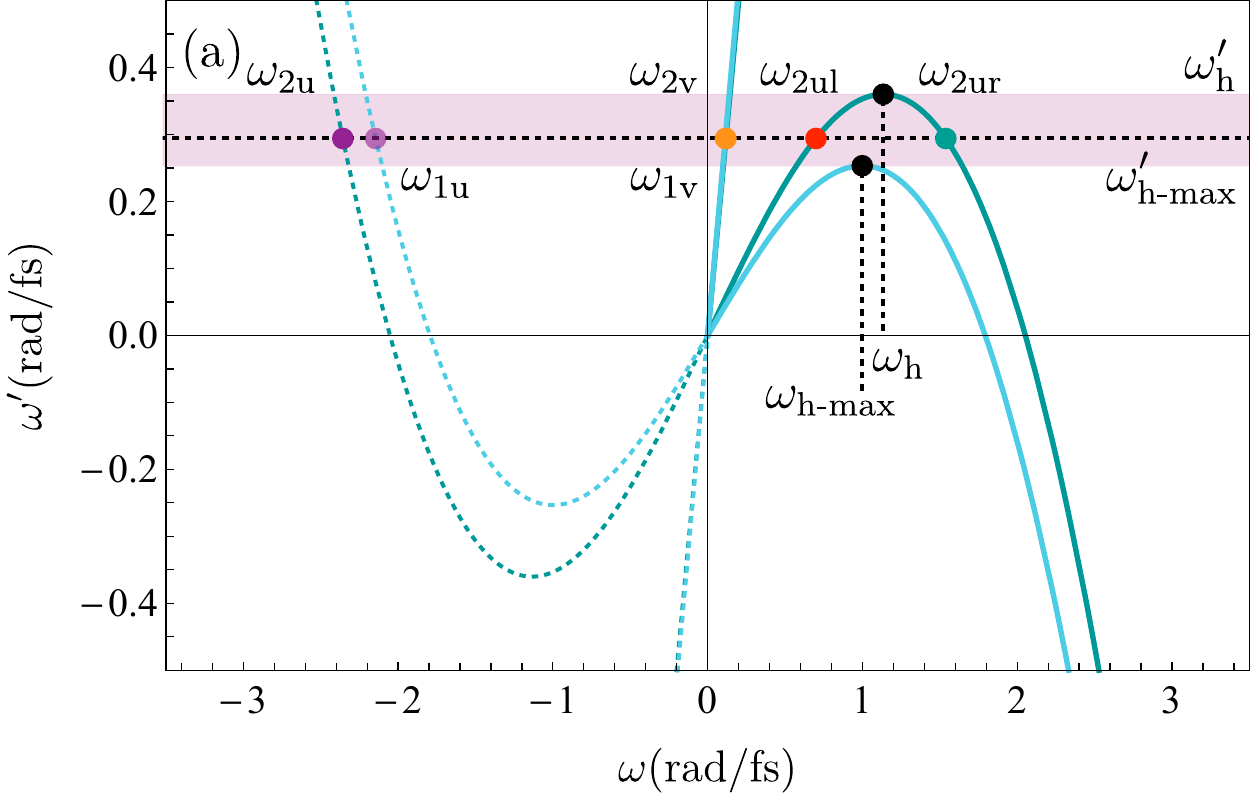}\hspace{1mm}
	\includegraphics[width=0.35\textwidth]{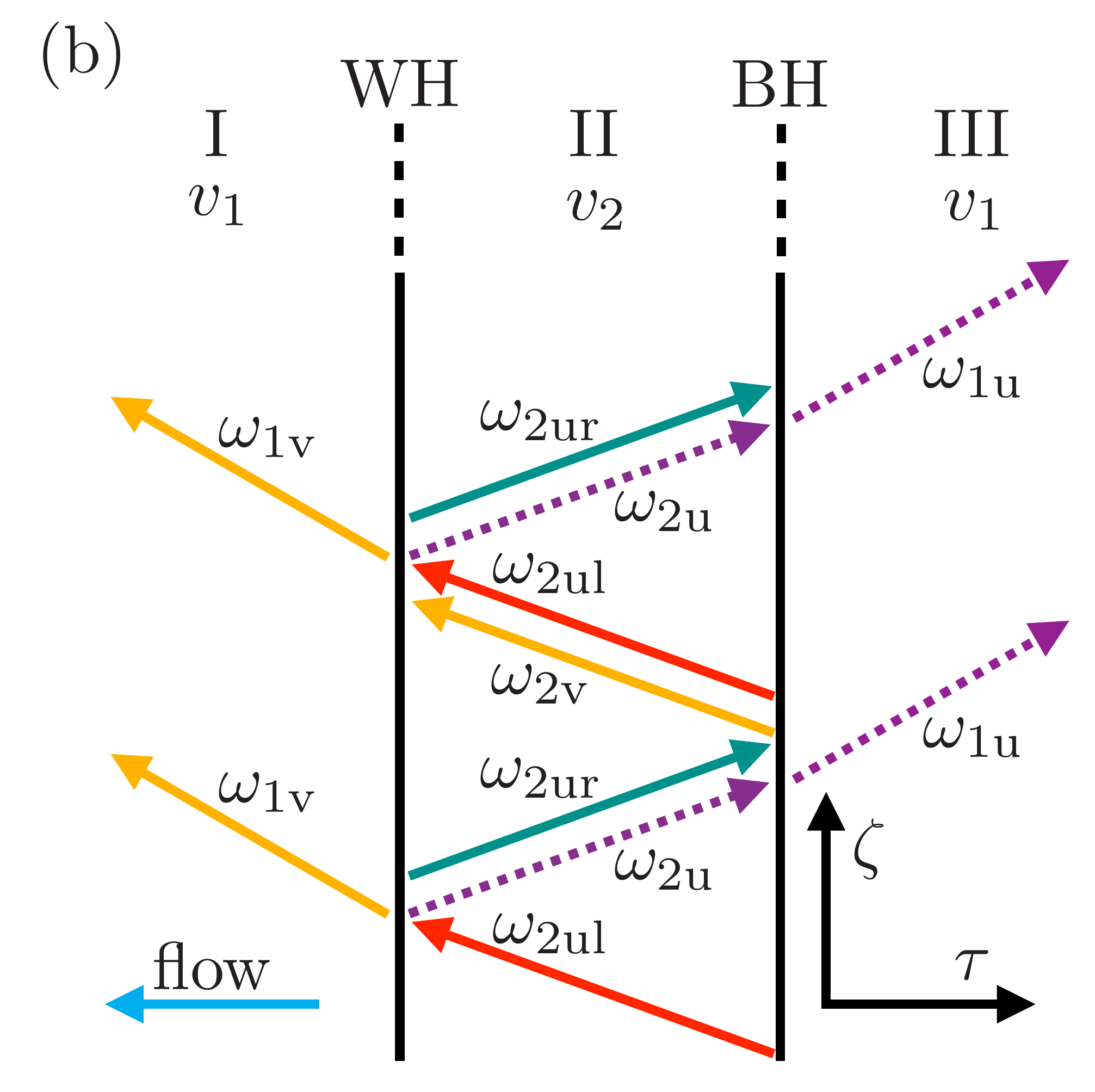}
	\caption{(a) Dispersion relation \eqref{disperoptica} in the comoving frame for the subluminal regions with $\delta n_\text{max}$ (blue) and for the superluminal region with $\delta n=0$ (green). The shadowed stripe marks the interval of comoving frequencies for the OBHL configuration. The conservation of comoving frequency gives two modes $\omega_{\text{1u}}$ and $\omega_{\text{1v}}$ in the subluminal regions and four $\omega_{\text{2u}}$, $\omega_{\text{2v}}$, $\omega_{\text{2ur}}$, and $\omega_{\text{2ul}}$ in the superluminal region. The parameters are $\omega'=0.294\text{rad/fs}$, $\Omega_0=1.861\text{rad/fs}$, $n_{g0}=1.488$, and $\delta n_\text{max}=0.1$. (b) Diagram of the evolution of a trapped in-mode $\omega_{2\text{ul}}$ for real frequencies, the escaping mode $\omega_\text{1u}$ is the resonant Hawking radiation. Solid [dashed] lines refer to positive [negative] frequency modes.}
	\label{fig2disp}
\end{figure}

The OBHL configuration is valid only for $\omega'$ close to the horizon $\omega'_\text{h}$ and limited by $\delta n_\text{max}$, see Section \ref{sechorizons}. The dispersion relation \eqref{disperoptica} is shown in Fig. \ref{fig2disp}(a), where the shadowed region indicates the region of comoving frequencies $(\omega'_\text{h-max},\omega'_\text{h})$ trapped by the cavity. If a fluctuation is spontaneously generated in the cavity with frequency $\omega_{\text{2ul}}$, its comoving frequency $\omega'(\omega_{\text{2ul}})$ is conserved. This mode can be dispersed into four possible modes inside the cavity and two outside. These modes are  obtained by numerically solving Eq. \eqref{disperoptica} for $v_1$ and $v_2$  and  are shown in Fig. \ref{fig2disp}(a).

The frequency modes that solve Eq. \eqref{disperoptica} are labeled with a subindex `u' for counterpropagating or `v' for copropagating modes and `1' for those in the subluminal region $v_1$ or `2' in the superluminal one $v_2$. In Fig. \ref{fig2disp}(a), the copropagating branch for the relevant frequencies is almost a straight line, reflecting that these modes are not dispersed, while the counterpropagating branch is a highly dispersive curve. For $\omega<0$, the solutions labeled as $\omega_\text{1u}$ and $\omega_\text{2u}$ are negative-frequency or negative-norm modes \cite{Robertson2011}. For $\omega>0$ and $v_1$ there is only one solution in the copropagating branch $\omega_\text{1v}$, for $v_2$ there are three modes, one in the copropagating branch $\omega_\text{2v}$ and two in the counterpropagating one $\omega_\text{2ul}$ and $\omega_\text{2ur}$. The labels in the last two solutions specify if the modes move to the left (l) or right (r) in the comoving frame. Modes $\omega_\text{2ul}$ and $\omega_\text{2ur}$ can be confined in the cavity \cite{Leonhardt2007}, as they fulfill the resonance condition, see Fig. \ref{fig2disp}(b) and Section \ref{sec5compa}.

\subsection{Direction of travel}\label{sec3Btravel}
\begin{figure}
	\centering
	\includegraphics[width=0.49\textwidth]{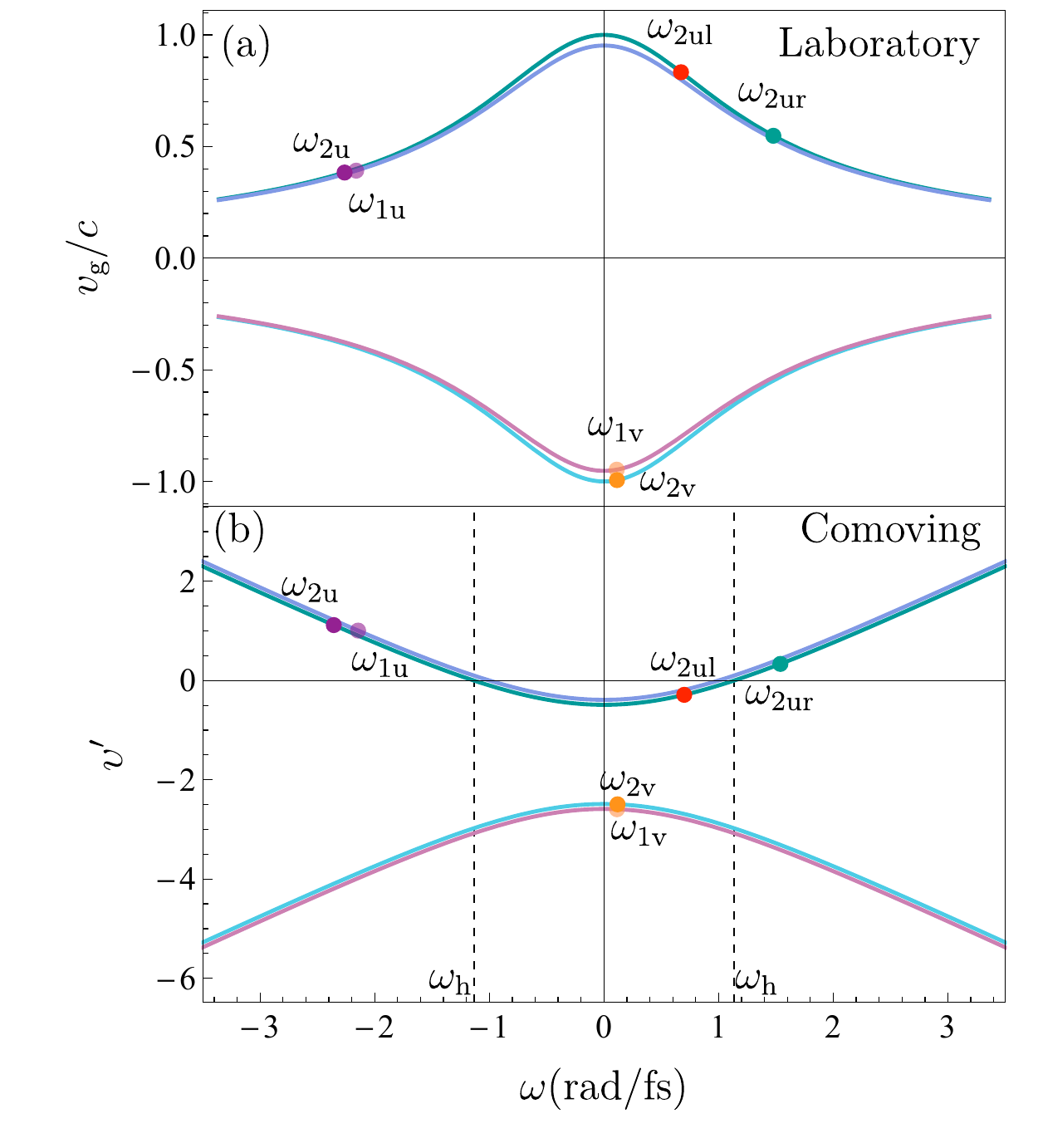}
	\caption{Group velocity in the laboratory frame(a) and dimensionless velocity in the comoving frame (b) for the 2u (green), 1u (purple), 1v (pink), 2v (blue) modes. The horizons $\pm \omega_\text{h}$ are marked with dashed lines and the solutions from Fig. \ref{fig2disp} with colored circles.}
	\label{fig3vels}
\end{figure}

In the laboratory frame, the direction of travel is given by the sign of the group velocity $v_g$, obtained from the dispersion relation \eqref{disperoptica} as:
\begin{equation}
v_g(\omega,\delta n)=\frac{\partial\omega}{\partial \beta}=\frac{c}{\delta n+
	\displaystyle\frac{2\omega^2+\Omega_0^2}{\omega^2+\Omega_0^2}\displaystyle \sqrt{1+\frac{\omega^2}{\Omega_0^2}}}.
\end{equation}
The counterpropagating modes (u) move with positive velocity (to the right in the laboratory frame) and the copropagating modes (v) move with negative velocity
\begin{equation}
v_\text{u}(\omega,\delta n)=v_g(\omega,\delta n),
\quad
v_\text{v}(\omega,\delta n)=-v_g(\omega,\delta n).
\end{equation}
We can describe the modes in the subluminal region with $\delta n=\delta n_\text{max}$ and in the superluminal region with $\delta n=0$. The normalized group velocity in the laboratory frame $v_g/c$ is shown for all modes in Fig. \ref{fig3vels}(a). As the dispersion is normal, the velocity diminishes when the frequency increases.

The dimensionless velocity in the comoving frame $v'$ is given by
\begin{equation}
v'(\omega,\delta n)=-\frac{\partial\omega'}{\partial \omega}=-n_{g0}\pm\left(\delta n+\displaystyle \frac{2\omega^2+\Omega_0^2}{\omega^2+\Omega_0^2}\sqrt{1+\frac{\omega^2}{\Omega_0^2}}\right). 
\end{equation}
The minus sign in front of the derivative appears because this dimensionless velocity is calculated with respect to two time-coordinates. For the counterpropagating and copropagating modes we have
\begin{equation}
v'_\text{u}= -n_{g0}+\frac{c}{v_g(\omega,\delta n)},\quad v'_\text{v}=-n_{g0}-\frac{c}{v_g(\omega,\delta n)}.
\end{equation}
setting $\delta n =\delta n_\text{max}$ for the subluminal regions and $\delta n = 0$ for the superluminal region, respectively. These dimensionless velocities are shown in Fig. \ref{fig3vels}(b). The copropagating (v) modes are negative for any frequency $\omega$ and they move in the same direction as the flow (left). The counterpropagating (u) modes move opposite to the flow, except for the low frequencies, where they invert its direction of travel; the limiting frequencies are the horizons $\pm \omega_\text{h}$ and are marked with vertical dashed lines in Fig \ref{fig3vels}(b). Next, we will see the advantages of defining the horizon as a frequency in dispersive systems.

In the sonic case, only the counterpropagating curve in the supersonic region 2u inverts its direction of travel \cite{2018Bermudez}. In the optical case, both regions 1u and 2u can invert it because the dimensionless velocity fulfills $\delta n \ll n_{g0}$  in both regions for low frequencies. However, for energies close to the horizon---the ones we are interested in---only 2u modes exist (conserve comoving energy) in that region, making the situation akin to the sonic analog \cite{2018Bermudez}.

\subsection{Horizons}\label{sechorizons}
We have previously defined a horizon as a point in space that separates the superluminal and subluminal regions of spacetime for the modes of the field, as in Fig. \ref{fig1diagram}. We can easily find its location if we match the velocity of the fluctuation with the velocity of the moving medium. In a dispersionless case, as in astrophysics, the horizon is a single spatial point for all frequencies of the quantum field. In a dispersive case, as is usual in analog systems, the horizon is fuzzy in space. This implies that each frequency has a different blocking point in space, but it is still uniquely defined in frequency.

For this reason, in dispersive systems it is more convenient to define the horizon as the frequency $\omega_\text{h}$ where the velocity of counterpropagating modes in the superluminal region $v'_\text{2u}$ is zero, that is, they are blocked. The definition is then

\begin{equation}
v'_\text{u}(\omega)|_{\omega=\omega_\text{h}}=0.
\end{equation}
Solving this equation, we find
\begin{equation}\label{horizontes}
\omega_\text{h}=\pm\frac{\Omega_0}{2\sqrt{2}}\sqrt{v^2-4\pm v\sqrt{v^2+8}},
\end{equation}
with $v=v_2$. For example, we obtain $\omega_\text{h-max}$ using $\delta n_\text{max}$. The corresponding values for the comoving frequency are $\omega'_\text{h}$ and $\omega'_\text{h-max}$, respectively, as shown in Fig. \ref{fig2disp}(b). The general solution for the horizon for any $\delta n$ is obtained with $v=v'_\text{u}$, the counterpropagating velocity in the comoving frame. This equation is equivalent to Eq. (24) in Ref. \cite{Larre2012} and Eq. (3.4) in Ref. \cite{2018Bermudez}. We can find an exact solution for the simple quartic dispersion of our system.

\subsection{Transluminal velocity}
Now we consider the minimal velocity $v(\tau)$ to reach the horizon, that is, given $\omega'$, $n_{g0}$, and $\Omega_0$, what is the velocity for modes $\omega_\text{2ur}$ and $\omega_\text{2ul}$ to exist (be real) and be equal? We call this the transluminal velocity $v_{t}$. In addition, we define the transluminal refractive index as $\delta n_{t}=n_{g0}+v_{t}$. Following Ref. \cite{2018Bermudez}, $v_t$ can be found analytically using an auxiliary function $q(\omega',\Omega_0)$:
\begin{equation}
q=270\frac{\Omega_0^2}{\omega'^2}+729\frac{\Omega_0^4}{\omega'^4}+\frac{3^{3/2}\Omega_0}{\omega'}\Bigl(27\frac{\Omega_0^2}{\omega'^2}-4\Bigr)^{3/2}-2,
\end{equation}
obtaining
\begin{equation}
v_{t}=- \sqrt{1+\frac{\omega'^2}{3\Omega_0^2}\left(\frac{q^{1/3}}{2^{1/3}}-1\right)
	+\frac{2^{1/3}}{3q^{1/3}}\left(\frac{\omega'^2}{\Omega^2_0}+54\right)}.
\end{equation}
For example, to fit the infrared horizon in an optical fiber used in analog gravity experiments \cite{drori2019observation}, the values are $\omega'=0.294$ rad/fs and $\Omega_0=1.862$ rad/fs. Thus, we find that $v_{t}=-1.428$. Recalling that $n_{g0}=1.488$, then $\delta n_t=0.06<\delta n_\text{max}=0.1$, these frequencies are blocked by the horizons, remain trapped inside the cavity, and are amplified by the Hawking effect; only the resonant Hawking radiation escapes the cavity.

\section{Instabilities}\label{sec4insta}
Instabilities are inherent to the analysis of the dynamics of fluctuations generated by a moving medium \cite{Leonhardt2003}. Thus far, we have described these fluctuations using its normal modes, that is, plane waves of the form $\text{e}^{-i(\omega \tau+\omega' \zeta)}$, as in Eq. \eqref{econdaplana2}, where $\omega$ and $\omega'$ are the frequencies in the laboratory and comoving frames, respectively. These frequencies are usually considered real to describe a plane wave, but in general, they can be complex to describe amplification or attenuation processes \cite{Hydrodynamic}.

We are interested in the process that amplifies Hawking radiation produced as a fluctuation inside the cavity. We take $\omega=\omega_R+i\omega_I$ and $\omega'=\omega'_R+i\omega'_I$ with $\omega'_I(\omega)>0$, such that the fluctuation is square-integrable in the delay time $\tau$ and grows exponentially with the propagation time $\zeta$: these mode fluctuations are the instabilities. On the other hand, modes with $\omega'_I<0$ are known as damped or stable \cite{gallaire2017fluid} and modes with $\omega'_I=0$ are called neutral, like those in the previous section. Neutral modes have been used to describe a classical version of the field fluctuation $\phi$ in the OBHL and to obtain its evolution and amplification \cite{Faccio2012,GaonaReyes2017}.

All previous works on instabilities are in fluid analogs, where the frequency is considered complex to describe the amplification \cite{zapata2011resonant,Macher2009pra, Coutant2010,Finazzi2010njp,Finazzi2015prl,2018Bermudez}. This is the first work of instabilities in the optical analog, where the comoving frequency is complex to account for the amplification in propagation time. The role of the wavenumbers in fluids is analogous to that of the laboratory frequencies in optics. By also considering them as complex, we allow for more flexibility to fulfill the conditions of instabilities than for resonances. In some cases, this leads to a larger number of instabilities than resonances. Even if there are several instabilities in a cavity, the one with the largest $\omega'_I$ quickly dominates the amplification process.

\subsection{Analytic method: OBHL with complex frequencies}
In this section, we apply the theory of instabilities to the OBHL. First, we need to generalize the previous solutions to complex values of $\omega$ and $\omega'$. Notice that Eq. \eqref{disperoptica} is a quartic equation in $\omega$ and its canonical form is
\begin{equation}
\omega^4+d\omega^3+e\omega^2+f\omega+g=0,
\end{equation}
with the following coefficients
\begin{equation}
d=0,\quad e=\Bigl(1-v^2\Bigr)\Omega_0^2,\quad f=-2v\omega' \Omega_0^2,\quad g=-\omega'^2\Omega_0^2.
\end{equation}
This type of equations with $d=0$ can be reduced to an auxiliary cubic equation
\begin{equation}
\omega^3+\frac{e}{2}\omega^2+\frac{e^2-4g}{16}\omega-\frac{f^2}{64}=0.
\end{equation}
The corresponding solutions $p_1$, $p_2$, and $p_3$ can be obtained in the usual way. The four solutions of the original equation can be recovered from them as:
\begin{subequations}\begin{align}
	\omega&=\sqrt{p_1}\pm(\sqrt{p_2}+\sqrt{p_3}),\\
	\omega&=-\sqrt{p_1}\pm(\sqrt{p_2}-\sqrt{p_3}).
	\end{align}\end{subequations}

\begin{figure}
	\centering
	\includegraphics[width=0.49\textwidth]{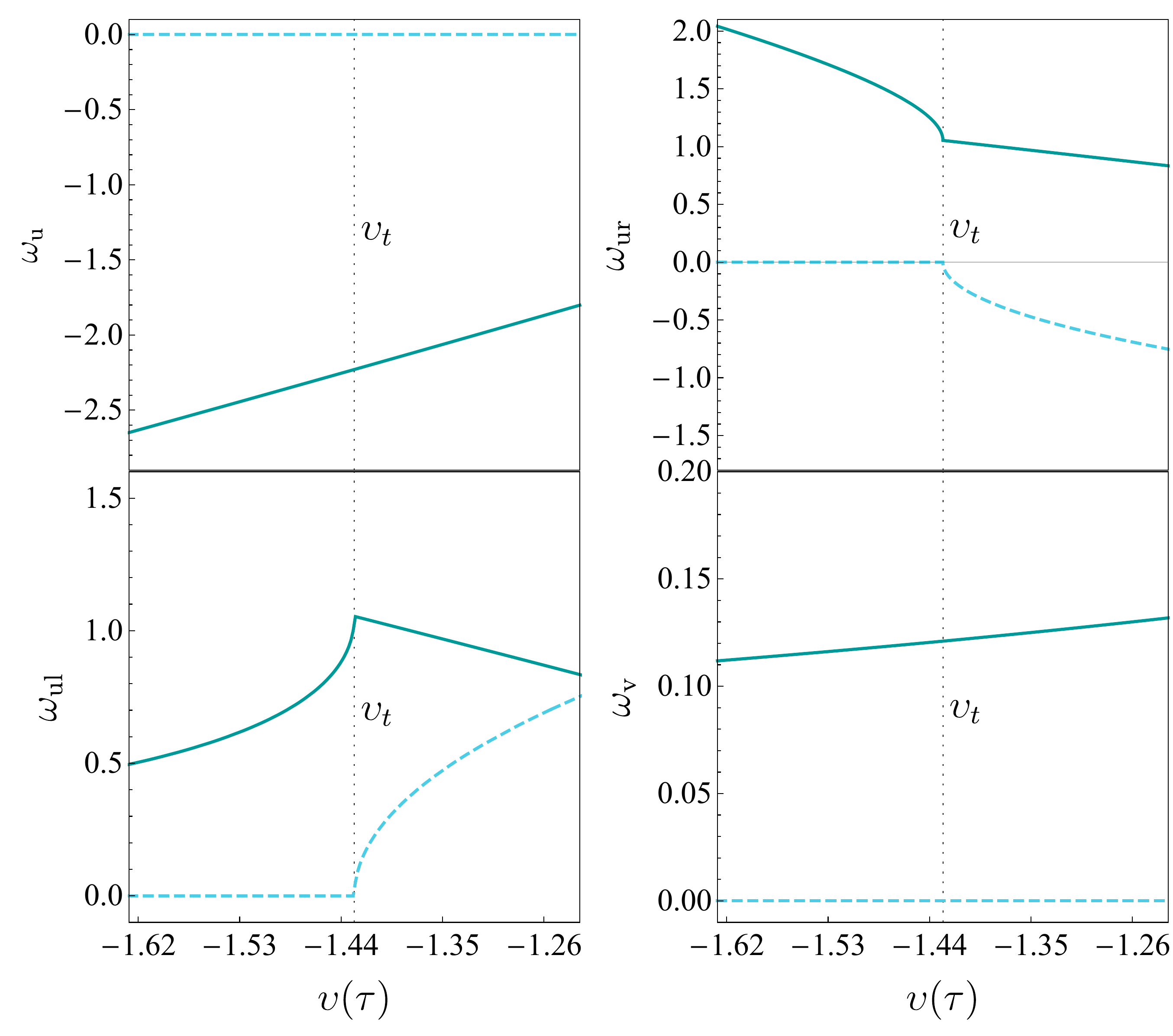}
	\caption{The four analytic solutions---real (green solid) and imaginary parts (blue dashed)---in terms of the fluid velocity $v$ for fixed values of $\omega'=0.294$rad/fs and $\Omega_0=1.861$rad/fs. The transluminal velocity $v_t$ (vertical dashed) is also shown. The solutions $\omega_{\text{u}}$ and $\omega_{\text{v}}$ are real for any $v$, but $\omega_\text{ur}$ and $\omega_\text{ul}$ are complex for $v>v_t$.}
	\label{fig4sols}
\end{figure}

These are the analytic solutions for the dispersion relation \eqref{disperoptica}. These expressions are cumbersome, but can be used for analytic calculations in symbolic software. In this way, we obtain all four solutions in each region of the OBHL. The behavior of the four complex solutions depends on the value of $v(\tau)$, as seen in Fig. \ref{fig4sols}. The transluminal velocity $v_t$ marks a change of behavior in the solutions, where $\omega_\text{ur}$ and $\omega_\text{ul}$ become complex.

\begin{figure}\centering
	\includegraphics[width=0.49\textwidth]{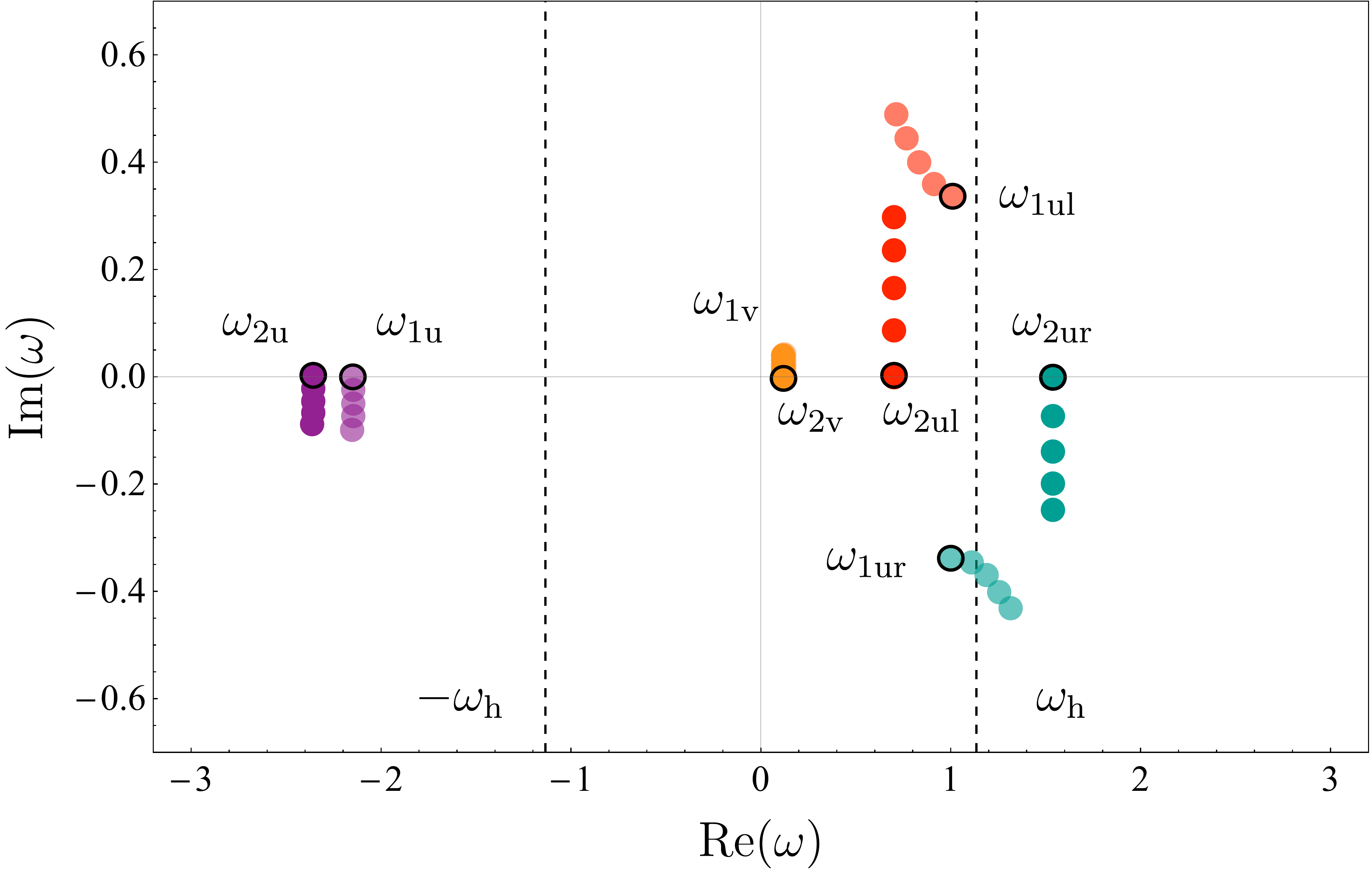}
	\caption{Analytic solutions in the complex space of $\omega$ after adding a small imaginary part $\omega'_I\ge 0$ to $\omega'$. These are the same six real solutions plus two new ones, completing four solutions for both regions. The solutions for $\omega'_I= 0$ are marked with black outline. Dashed lines in $\omega_\text{h}$ help us find the direction of travel for the 2u modes: $\omega_\text{2u}$, $\omega_\text{2ur}$, and $\omega_\text{2ul}$.}
	\label{fig5cspace}
\end{figure}

Let us return to the solutions analyzed in Section \ref{sec3obhl} and shown in Figs. \ref{fig2disp} and \ref{fig3vels}. The analytic solutions for the same conditions are obtained in the complex $\omega$ plane and shown as points with black outline in Fig. \ref{fig5cspace}. We obtain the same six results from the numerical method: four in the superluminal region and two in the subluminal one, but also two extra solutions in the subluminal region that are complex conjugate of each other and do not appear in the initial treatment of the OBHL. These are called evanescent modes \cite{Larre2012, Mathieu2017}, and although their norm is zero, it is important to include them in scattering calculations, as in Ref. \cite{Isoard2020}. These additional solutions describe exponentially growing or decreasing modes, as their $\tau$-dependence in Eq. \eqref{econdaplana2} is given by
\begin{equation}\label{complex}
\text{e}^{-i\omega \tau}=\text{e}^{-i(\omega_R+i\omega_I)\tau}=\text{e}^{\omega_I\tau}\text{e}^{-i\omega_R\tau}.
\end{equation}

\subsection{Qualitative and quantitative description}
We search for the instabilities of the OBHL: Modes with complex comoving frequency $\omega'=\omega'_R+i\omega'_I$ such that $\omega'_I>0$ and, in consequence, their amplitude increases exponentially with the propagation time
\begin{equation}\label{exps}
\text{e}^{-i\omega' \zeta}=\text{e}^{-i(\omega'_R+i\omega'_I)\zeta}=\text{e}^{\omega'_I\zeta}\text{e}^{-i\omega'_R\zeta}.
\end{equation}

We start by finding the conditions that the quantum fluctuation $\phi(\tau)$ from Eq. \eqref{econdaplana2} must satisfy in $\tau$. Considering Eq.  \eqref{complex} and dividing $\phi(\tau)$ in three regions I, II, and III, as shown in Fig. \ref{fig1diagram}(b):
\begin{subequations}\begin{align}
	\phi_\text{I}		&=\textbf{A}\cdot\text{e}^{-i{\boldsymbol\omega}_1\tau},\qquad\textbf{A}=(A_1,A_2,A_3,A_4),\\
	\phi_\text{II} 	&=\textbf{B}\cdot\text{e}^{-i{\boldsymbol\omega}_2\tau},\qquad\textbf{B}=(B_1,B_2,B_3,B_4),\\
	\phi_\text{III}	&=\textbf{C}\cdot\text{e}^{-i{\boldsymbol\omega}_1\tau},\qquad\textbf{C}=(C_1,C_2,C_3,C_4),
	\end{align}\end{subequations}
where
\begin{subequations}\begin{align}
	\text{e}^{-i{\boldsymbol\omega}_1\tau}&=(\text{e}^{-i\omega_{11}\tau},\text{e}^{-i\omega_{12}\tau},
	\text{e}^{-i\omega_{13}\tau},\text{e}^{-i\omega_{14}\tau}),\\
	\text{e}^{-i{\boldsymbol\omega}_2\tau}&=(\text{e}^{-i\omega_{21}\tau},\text{e}^{-i\omega_{22}\tau},
	\text{e}^{-i\omega_{23}\tau},\text{e}^{-i\omega_{24}\tau}).
	\end{align}\end{subequations}
A global mode of the cavity depends on $\textbf{A},\textbf{B},\textbf{C}$, and $\omega'$. The value of $\omega'$ fixes the four values ${\boldsymbol\omega}_1$ in the subluminal region and another four ${\boldsymbol\omega}_2$ in the superluminal one. The three vectors $\textbf{A},\textbf{B},\textbf{C}$ describe the mode amplitudes, each vector depends on four complex coefficients.

\begin{figure}\centering
	\includegraphics[width=0.48\textwidth]{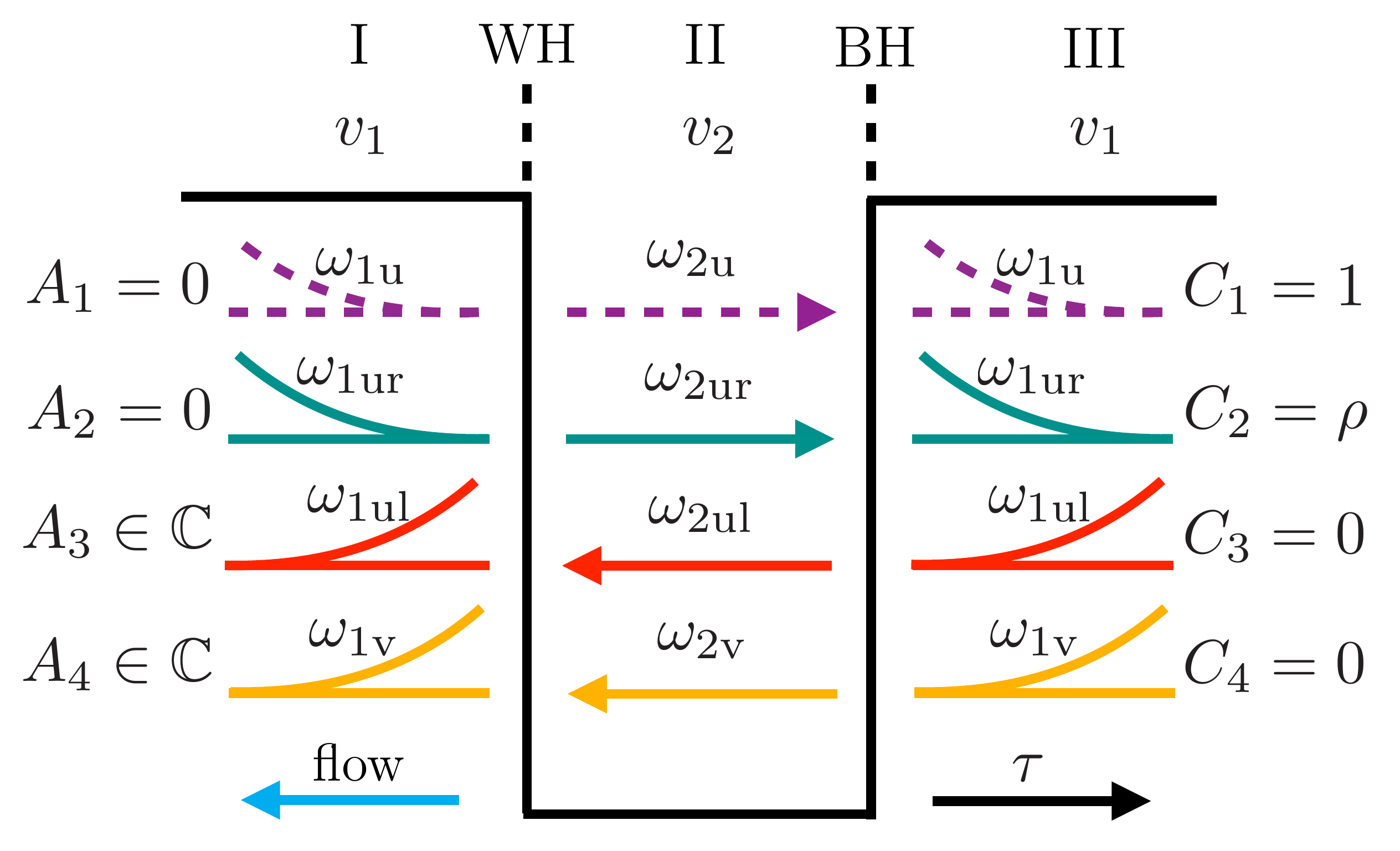}
	\caption{Mode diagram in $\tau$ for an OBHL with instabilities. The coefficients $A_j,C_j$ shown describe the most general normalizable solution for the field $\phi$. Solid [dashed] lines refer to positive- [negative-] frequency modes.}
	\label{fig6cavity}
\end{figure}

We consider only normalizable cavity states, that is, the modes that decay far away from the cavity by setting to zero the amplitude of the unbounded modes outside the cavity. From the values in Fig. \ref{fig5cspace} and taking into account the Eq. \eqref{complex}, we notice that $\text{Im}(\omega_\text{1ul})>0$, so it decays to the left ($\tau\rightarrow -\infty$), while $\text{Im}(\omega_\text{1ur})<0$ and it decays to the right ($\tau \rightarrow \infty$). When we add a small imaginary part $\omega'_I>0$ to $\omega'$, all $\omega$ values become complex too (colored points in Fig. \ref{fig5cspace}). Modes with Im($\omega)>0$ increase with $\tau$ and modes with Im($\omega)<0$ decrease with it, as depicted in Fig. \ref{fig6cavity}. We restrict the fluctuation $\phi(\tau)$ to be square-integrable and therefore set $A_1 = A_2 = C_3 = C_4 = 0$. The remaining coefficients $A_3,\ A_4,\ C_1$, and $C_2$ should be found to completely describe a cavity mode.

\subsection{Confined states}\label{secconfined}
We use the transfer matrix method to obtain the cavity modes \cite{2018Bermudez}. We can use this method under the step-index approximation because the velocity profile $v(\tau)$ is always constant, except at the interfaces, where the solution can be found by the continuity of the quantum field $\phi(\tau)$ and its first three derivatives there, given our fourth-order equation. To define the transfer matrices $M_1$ from region I to II and $M_2$ from region II to III, we use two auxiliary matrices $m_1$ and $m_2$, whose entries are the four amplitudes of the field and its derivatives ($\phi,\phi', \phi'',\phi'''$), that is,
\begin{align}
m_1=\begin{pmatrix}
1 & 1 & 1 & 1 \\
-i\omega_{11} &-i\omega_{12}  & -i\omega_{13} &-i\omega_{14} \\
-\omega_{11}^2 &- \omega_{12}^2 &-\omega_{13}^2  &-\omega_{14}^2 \\
i\omega_{11}^3&  i\omega_{12}^3&  i\omega_{13}^2 & i\omega_{14}^3
\end{pmatrix},\\
m_2=\begin{pmatrix}
1 & 1 & 1 & 1 \\
-i\omega_{21} &-i\omega_{22}  & -i\omega_{23} &-i\omega_{14} \\
-\omega_{21}^2 &- \omega_{22}^2 &-\omega_{23}^2  &-\omega_{24}^2 \\
i\omega_{21}^3&  i\omega_{22}^3&  i\omega_{23}^2 & i\omega_{24}^3
\end{pmatrix}.
\end{align}

The first subindex $i$ in $\omega_{ij}$ corresponds to the velocity $v_1$ or $v_2$, and the second one $j$ to the particular solution 1-4 always ordered as u, ur, ul, and v---the same order as in Fig. \ref{fig6cavity}. Then, the transfer matrices are simply
\begin{equation}
M_1=m_2^{-1} m_1, \qquad M_2=m_1^{-1} m_2=M_1^{-1}.
\end{equation}
It is also convenient to define matrices that propagate the solution between the two interfaces of the cavity. These are
\begin{align}
P_\text{L}&=\text{diag}(\text{e}^{i\omega_{21}\tau_\text{c}},\text{e}^{i\omega_{22}\tau_\text{c}},\text{e}^{i\omega_{23}\tau_\text{c}},\text{e}^{i\omega_{24}\tau_\text{c}}),\\
P_\text{R}&=\text{diag}(\text{e}^{-i\omega_{21}\tau_\text{c}},\text{e}^{-i\omega_{22}\tau_\text{c}},\text{e}^{-i\omega_{23}\tau_\text{c}},\text{e}^{-i\omega_{24}\tau_\text{c}}).
\end{align}
Finally, the transfer matrix $M$ from region III to I is
\begin{equation}\label{matrizM}
M=M_1P_\text{L}M_2=m_2^{-1}m_1P_\text{L}m_1^{-1}m_2.
\end{equation}
With $M$ we can obtain the coefficients $\textbf{A}$ in terms of $\textbf{C}$ from
\begin{equation}
\phi_\text{I}=M\phi_\text{III}.
\end{equation} 
Similarly, we can obtain  $\textbf{C}$ in terms of $\textbf{A}$ using $M^{-1}$.

Let us state the problem: starting from a given comoving frequency $\omega'$ and four coefficients $\textbf{C}$ [$\textbf{A}$], we can determine the eight frequencies ${\boldsymbol\omega}_1(\omega')$ and ${\boldsymbol\omega}_2(\omega')$ and the other eight coefficients $\textbf{A}$ and $\textbf{B}$ [$\textbf{B}$ and $\textbf{C}$]. Remember that all these quantities are complex.

With the conditions in Fig. \ref{fig6cavity}, we define a spontaneous lasing mode or instability as a mode where the fluctuation $\phi(\tau)$ is square-integrable and the imaginary part of its comoving frequency $\omega'$ is positive. For a given value $\omega'$, we can obtain analytically the frequencies ${\boldsymbol\omega}_1$ and ${\boldsymbol\omega}_2$ and one of the coefficients, but the equation for the second one is transcendental and needs to be solved numerically. We can find the instabilities or lasing modes $\omega'_\ell$ by varying the two parameters that describe the cavity: the pulse separation $\tau_\text{c}$---like ``length''---and the optical contrast $\delta n_\text{max}$---like ``height''. In Table \ref{table1} we show the number of instabilities $N_\text{ins}$ found for cavities with parameters $\tau_\text{c}=\{6,\ 13,\ 20\}$fs and $\delta n_{\text{max}}=\{0.01, \ 0.05\}$.

\begin{figure*}
	\centering
	\includegraphics[width=0.95\textwidth]{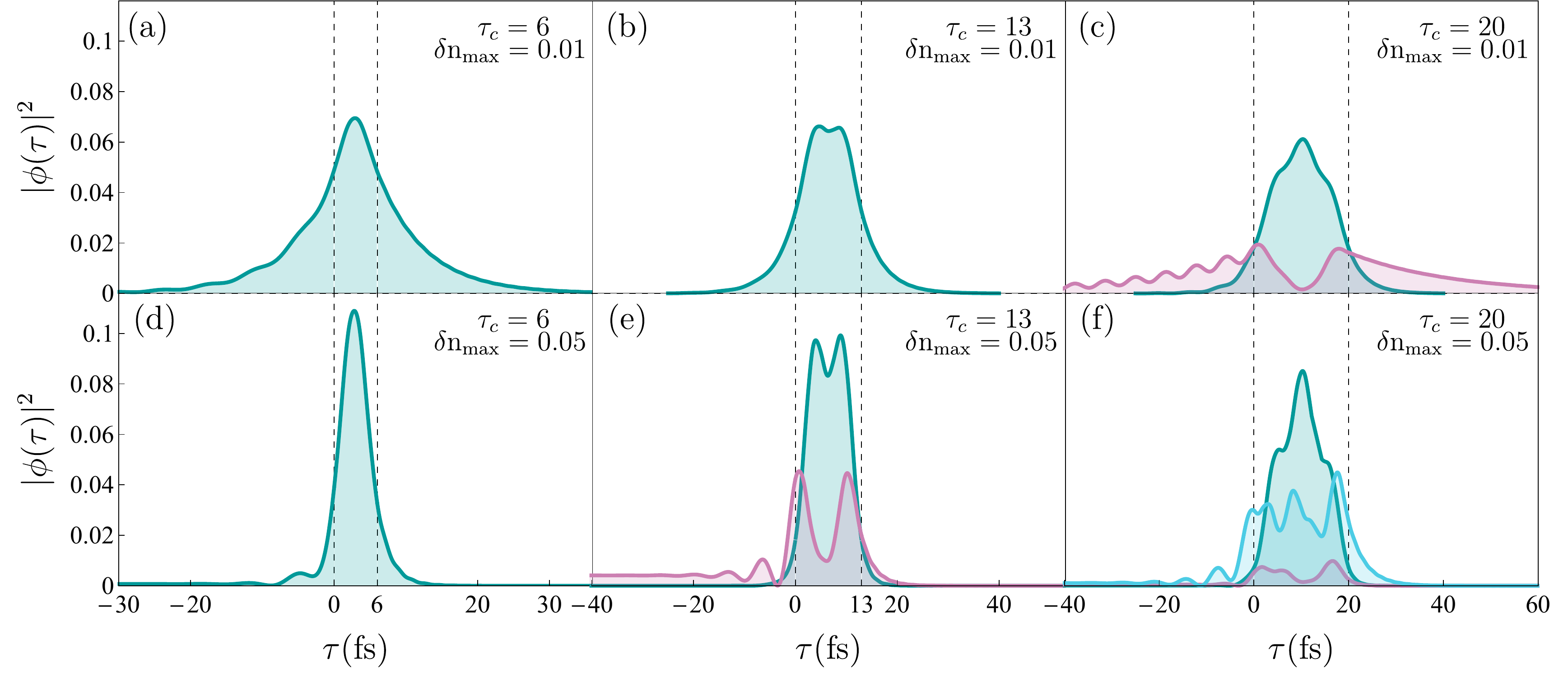}
	\caption{Probability density $|\phi(\tau)|^2$ of the instabilities for the cavities in Table \ref{table1}.In these examples, the ground state (green) always exists, the first excited state (pink) in three cases, and the second excited state (blue) in one. The vertical dashed lines mark the cavity.}
	\label{fig7prob}
\end{figure*}

We also calculated the confinement ratio $P_\text{c}$ that specifies how much of the probability distribution of the quantum fluctuation $\phi$ is inside the cavity:
\begin{equation}
P_{\text{c}}=\int_0^{\tau_\text{c}} d\tau \ |\phi(\tau)|^2,
\end{equation}
where $|\phi|^2$ is normalized from $-\infty$ to $\infty$. In Fig. \ref{fig7prob} we show the probability density $|\phi|^2$ for the instabilities in Table \ref{table1}. 

In all the solutions found, $P_\text{c}< 1$ because a part of the field can leak out of the cavity before returning, but also because another part can escape beyond the horizons. This is the typical behavior of resonant Hawking radiation and is similar to the findings in the acoustic system \cite{2018Bermudez,curtis2019evanescent,Coutant2019}. From Table \ref{table1} and Fig. \ref{fig7prob}, we see that the confinement of the ground state increases with the size of the cavity, that is, increasing $\tau_\text{c}$ or $\delta n_\text{max}$.

\begin{table}
	\centering
	\begin{tabular}{|c|c|c|c|c|c|}
		\hline
		$\hspace{-0.mm}\tau_\text{c}\text{(fs)}\hspace{-0.mm}$& $\hspace{-0.mm}\delta n_\text{max}\hspace{-0.mm}$& $\hspace{-0.mm}N_\text{ins}\hspace{-0.mm}$&$\omega_\ell '$(rad/fs)& $\hspace{-0.mm}P_\text{c}(\%)\hspace{-0.mm}$& $\hspace{-0.mm}x_\text{a}\text{(m)}\hspace{0.mm}$\\
		\hline
		\multirow{2}{*}{6} & 0.01 & $1$&$0.351+i 1.64\cdot10^{-6}$&36&9.23\\ \cline{2-6}
		& 0.05 &1&$0.333+i6.12\cdot 10^{-5}$& 47&0.25\\ \cline{1-6}
		\multirow{3}{*}{13} & \multirow{1}{*}{0.01}& \multirow{1}{*}{1}& $0.355+i9.60\cdot10^{-7}$ &74&1.32\\ \cline{2-6}
		& \multirow{2}{*}{0.05}& \multirow{2}{*}{2}& $0.350+i1.45\cdot10^{-5}$&95&0.09\\ \cline{4-6}
		&&&$0.319+i4.87\cdot10^{-5}$&36&0.31\\ \cline{1-6}
		\multirow{5}{*}{20} & \multirow{2}{*}{0.01}& \multirow{2}{*}{2}& $0.356+i4.78\cdot10^{-7}$ &87&2.58\\ \cline{4-6}
		&&&$0.349+i5.08\cdot10^{-7}$&21&2.52\\ \cline{2-6}
		& \multirow{3}{*}{0.05}& \multirow{3}{*}{3}&$0.354+i5.02\cdot10^{-6}$ &96&0.26\\ \cline{4-6}
		&&&$0.337+i2.58\cdot10^{-5}$&11&0.06\\ \cline{4-6}
		&&&$0.312+i1.16\cdot10^{-5}$&60&0.12\\ \cline{1-6}
	\end{tabular}
	\caption{Parameters of instabilities in six different cavities with $\tau_\text{c}$ and $\delta n_\text{max}$.}
	\label{table1}
\end{table}

In this way, we can find a discrete set of instabilities $\omega'_\ell$ for a given geometry of the cavity $\tau_\text{c}$ and $\delta n_\text{max}$. The comoving frequencies $\omega'_\ell$ and confinement ratios $P_c$ for six cavities are shown in Table \ref{table1}, and the probability densities $|\phi|^2$ in Fig. \ref{fig7prob}. The number of instabilities $N_\text{ins}$ increases with $\tau_\text{c}$ and $\delta n_{\text{max}}$ in those examples. Is it possible to know how many instabilities are there given the parameters of the cavity? In the following section, we compare the solutions found by the theory of instabilities with the resonances of the plane-wave model used in Section \ref{sec3obhl} and show that it is possible.

\section{Instabilities vs resonances}\label{sec5compa}
In the previous section, we showed how to find the instabilities of an OBHL for any geometry of the cavity ($\tau_\text{c},\delta n_\text{max}$). Now, we use some simple models to get a better physical picture of the instabilities $\omega(\omega'_\ell)$ with complex frequency $\omega'_\ell=\omega'_R+i\omega'_I$ and the resonances in the plane-wave model with real frequency $\omega(\omega'_R)$ used in Section \ref{sec3obhl}, that is, by taking the limit $\omega'_I\rightarrow 0$.

According to Section \ref{sec3Btravel}, the modes trapped in the cavity are $\omega_\text{2ur}$ and $\omega_\text{2ul}$. Each time one of these modes reaches a horizon, it is amplified through the analog Hawking effect. Their evolution still conserves the norm due to the existence of negative-norm modes (dashed lines in Fig. \ref{fig2disp}). After each cycle, a mode $\omega_{\text{1u}}$ leaves the cavity and its amplitude increases. This is the so-called plane-wave model and it can verify if the instabilities fulfill a resonance condition \cite{Leonhardt2007,GaonaReyes2017}.

\subsection{Resonance condition}\label{sec5Ares}
A resonance is produced when trapped radiation has a phase difference of a multiple of $2\pi$ after a periodic travel \cite{Leonhardt2007,2018Bermudez}. We must consider the two reflections in a period, one for each horizon and each one producing a phase change of $\pi/2$. Then, the phase difference of trapped modes $\omega_\text{2ul}$ and $\omega_\text{2ur}$ should fulfill
\begin{equation}\label{match}
\tau_\text{c} \Delta \omega=(2n+1)\pi, \quad n=0,1,2,\dots .
\end{equation}
The first resonance ($n=0$) at $\tau_\text{c} \Delta \omega=\pi$ should be close to the ground state instability, the second resonance ($n=1$) to the first excited state instability, and so on.

\begin{figure}\centering
	\includegraphics[width=0.48\textwidth]{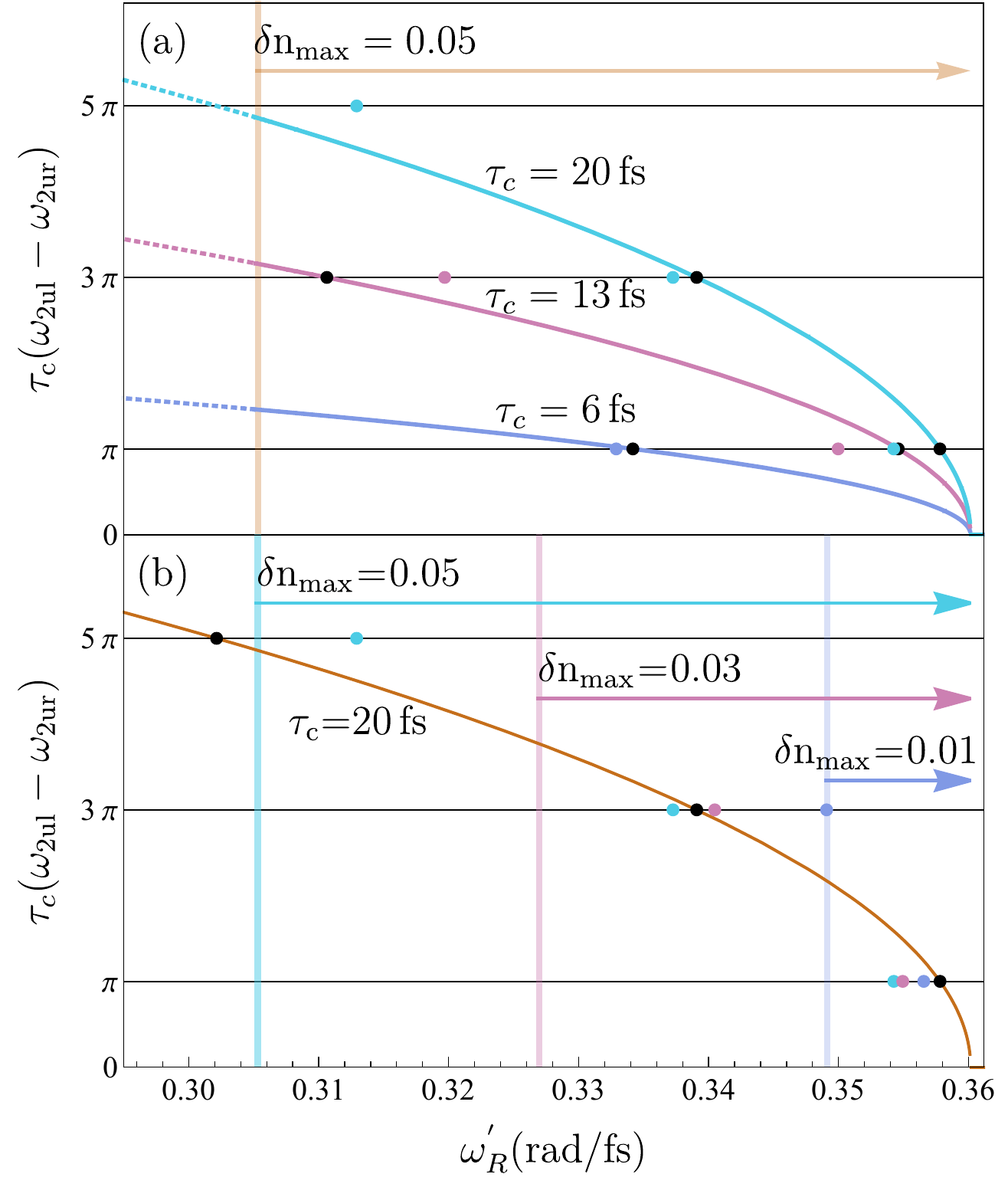}
	\caption{Phase difference $\tau_\text{c}\Delta \omega $ for (a) varying $\tau_\text{c}$ and fixed $\delta n_{\text{max}}=0.05$ and (b) fixed $\tau_\text{c}=20$fs and varying $\delta n_{\text{max}}$, marked with vertical dashed lines; resonances should be on the right side of these lines. The phase differences for $\pi$, $3\pi$ and $5\pi$ for the resonances are marked with black points and the instabilities with colored points.}
	\label{fig8diff}  
\end{figure}

We analyze the resonance condition from two points of view: by varying $\tau_\text{c}$ while keeping $\delta n_\text{max}$ fixed and by keeping $\tau_\text{c}$ fixed while varying  $\delta n_\text{max}$. Both solutions are shown in Fig. \ref{fig8diff}, the black points are the resonances resulting from the phase-matching condition in Eq. \eqref{match}, shown as a color curve. The colored points are the instabilities obtained in Section \ref{sec4insta} for $\tau_\text{c}=\{6,\ 13,\ 20\}$fs and  $\delta n_\text{max}=0.05$ in (a) and $\tau_\text{c}=20$ fs and $\delta n_\text{max}=\{0.01,\ 0.03,\ 0.05\}$ in (b).

In most cases, the frequencies $\omega'_R$ predicted for the resonances are close to those for the instabilities, their values differ less than $7\%$. Their discrepancy can be explained because the resonance condition \eqref{match} does not contain information about the ``height'' of the cavity $\delta n_\text{max}$. The only information imposed in the case of resonances is the minimum value $\delta n_\text{max}$ to trap the mode $\omega_R'$, given by the condition $\delta n_{\text{max}}>\delta n_{t}$, as shown in Fig. \ref{fig8diff}(b). The resonance condition in Eq. \eqref{match} changes for non-flat velocity profiles \cite{GaonaReyes2017,michel2013saturation}.

\begin{figure}\centering
	\includegraphics[width=0.48\textwidth]{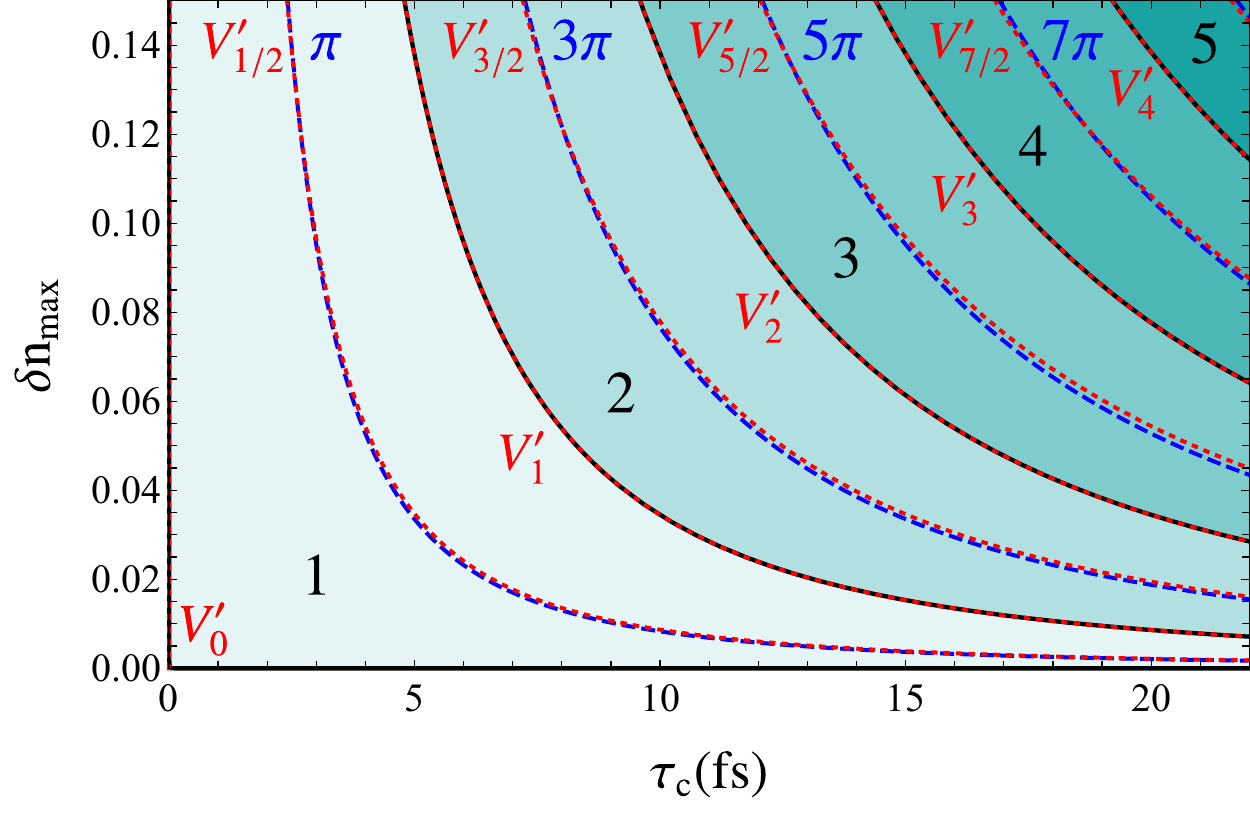}
	\caption{Number of instabilities and resonances in the parameter space of the geometry of the cavity ($\tau_\text{c},\delta n_\text{max}$). The colored regions separated by a black line show the number of instabilities $N_\text{ins}$ marked with large black numbers. The blue dashed lines mark the regions where a new resonance appears ($\pi,3\pi,5\pi,7\pi$). The red dotted lines come from the phenomenological model for the comoving frequency $V'$, matching the instabilities and resonances.}
	\label{fig9num}
\end{figure}

The number of resonances in the plane-wave model for the parameter space of the cavity geometry $(\tau_\text{c},\delta n_{\text{max}})$ are shown in Fig. \ref{fig9num}. The blue dashed lines show the parameters for a new resonance at $\pi,3\pi, 5\pi,7\pi$. The number of resonances at a certain point in the parameter space is given by the number of blue lines crossed from the zero axis (either $\tau_c$ or $\delta n_\text{max}$). We can compare it with the number of instabilities $N_\text{ins}$ found for several cases including those from Table \ref{table1} and Fig. \ref{fig7prob}. For example, in Fig. \ref{fig8diff}(b) for $\delta n=0.01$, there is only one resonance (for the $\pi$-value), whereas there is an extra instability for the $3\pi$-value at the limiting value of $\omega'_R$. The resonances approximate the number of solutions but they do not capture all the physics, for example, there are several geometries in Fig. \ref{fig9num} where there is one more instability than resonances. This is expected as there is more flexibility for an instability to be square-integrable because its parameters are complex $\omega,\omega'\in\mathbb{C}$. For this reason, it is convenient to study the instability condition.

\subsection{Instability condition}
\begin{figure}\centering
	\includegraphics[width=0.48\textwidth]{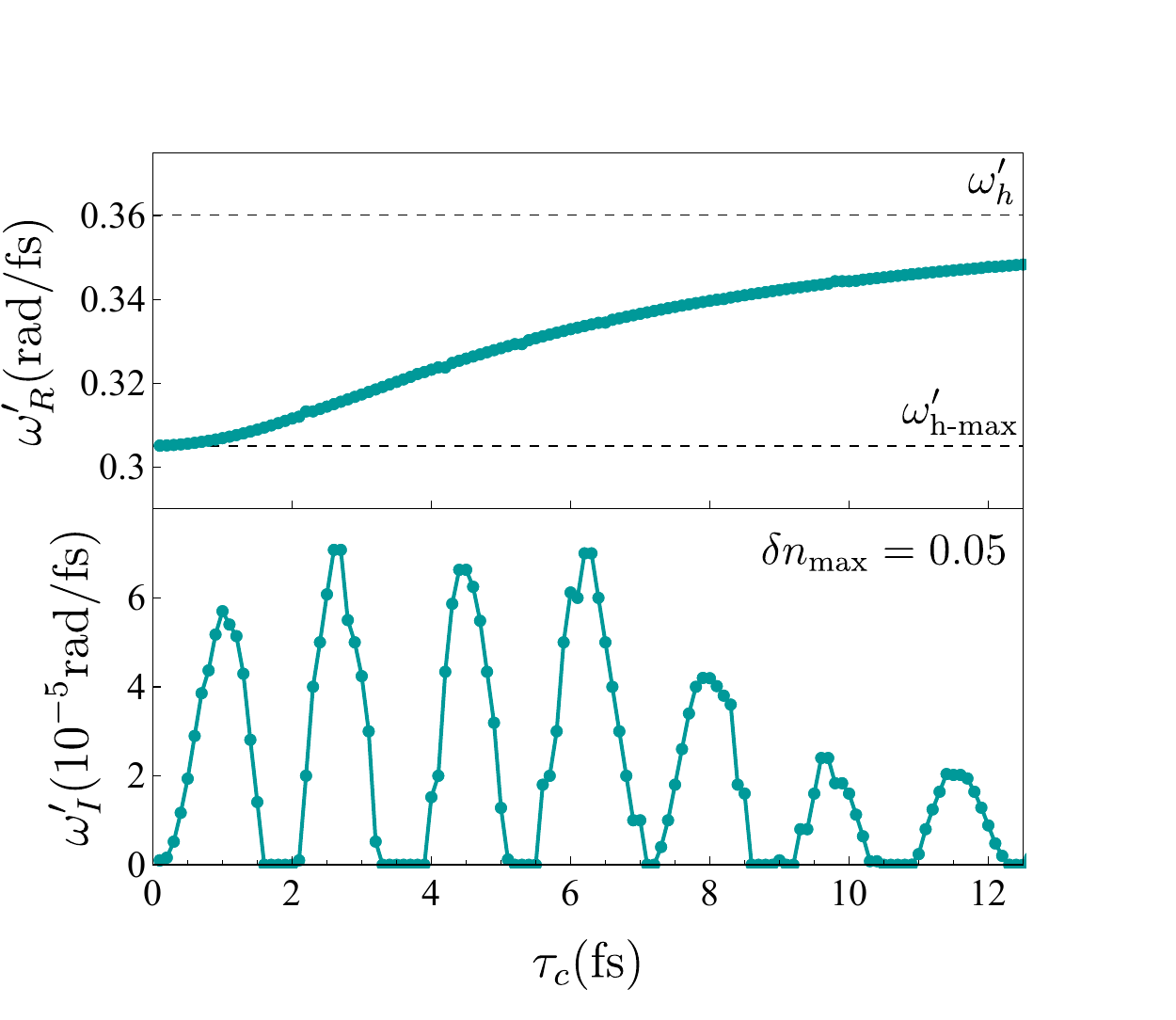}
	\caption{The instability condition allow us to follow $\omega'_R$ and $\omega'_I$ of the ground state instability for $\delta n_\text{max}=0.05$ and varying $\tau_c$. At $\tau_c\rightarrow 0$, the limiting values are $\omega'_R\rightarrow\omega'_\text{h-max}$ and $\omega'_I\rightarrow 0$.}
	\label{fig10reim}
\end{figure}

From the previous section, and in particular from the results in Fig. \ref{fig8diff}(a), we can see that new resonances appear when the dimensions of the cavity ($\tau_c$ and $\delta n_\text{max}$) increase such that a new resonance condition is met. This is initially fulfilled at the lowest allowed comoving frequency. The range of allowed comoving frequencies is the pink region in Fig. \ref{fig2disp}(a) and its lowest value is $\omega'_\text{h-max}$, obtained from Eq. \eqref{horizontes}. Following Ref. \cite{michel2013saturation}, we can also study the birth of an instability by increasing the length of the cavity ($\tau_c$ in our case) for the lowest allowed frequency. This is the instability condition. In the case shown in Fig. \ref{fig10reim} for fixed $\delta n_\text{max}$ and varying $\tau_c$, the ground state instability reaches the limits $\omega'_R\rightarrow \omega'_\text{h-max}$ and $\omega'_I\rightarrow 0$ for $\tau_c\rightarrow 0$.

We can solve the system of equations in Section \ref{secconfined} to obtain the confined states for the $\omega'_R\rightarrow \omega'_\text{h-max}$ with $\omega'_I\rightarrow 0$. This procedure produces the values of $\tau_c$ where a new instability appears for a given $\delta n_\text{max}$. We show the results in Fig. \ref{fig9num}. In particular, we obtained that even the smallest cavities have at least one instability. The six cases in Fig. \ref{fig7prob} match with the expected results from this model.

Fitting these numerical solutions, we obtain a phenomenological rule for the number of instabilities $N_\text{ins}$, given by
\begin{equation}\label{mA}
N_\text{ins}= 1 + \left\lfloor \frac{\tau_c\sqrt{\delta n_\text{max}}}{S}\right\rfloor,
\end{equation}
where $\lfloor\cdot\rfloor$ is the floor function and $S\simeq 1.867$ fs is an effective area (with delay units) for the cavity to fit an additional instability. In Fig. \ref{fig9num}, the number of instabilities $N_\text{ins}$ is marked with black numbers and the regions divided by black solid lines and colored with different shades of green. We confirm that for some parameters the number of instabilities and resonances agree, but also regions where there is one more instability than resonances, as we saw from the numerical solutions from Fig. \ref{fig7prob}. In fact, the resonance condition is in the middle of the corresponding instability region.

\subsection{Phenomenological model}
We introduce a phenomenological model to characterize the number of instabilities or resonant modes in a cavity. This model is taken directly from fiber optics, where it is used to calculate the number of modes supported by a fiber \cite{Agrawal2013}. In fiber optics, the cavity is constituted by the transversal dimension of the fiber, sometimes limited by a core. This model is described by the parameter $V$ or normalized frequency (although it is a dimensionless quantity), given by
\begin{equation}
V=k_0 a \sqrt{n_1^2-n_2^2},
\end{equation}
where $k_0$ is the wavenumber of the radiation, $a$ is the core radius, $n_1$ and $n_2$ are the two refractive indices that make the optical contrast of the cavity.

Performing the changes of variables to match our geometry, in which the radiation is trapped not in the transversal spatial dimension as in a fiber, but in the longitudinal temporal direction or delay, we can define the equivalent normalized comoving frequency or $V'$ parameter is
\begin{equation}\label{pheno}
V'=\omega'_0\tau_\text{c}\sqrt{2 n_0\delta n_\text{max}},
\end{equation}
where $\omega_0'=\omega'(\omega_h)$ and $n_0=n(\omega_h)$. This parameter is again dimensionless and depends on the geometry and contrast of the cavity.

The adjusted values of $V'$ are limited by red dashed lines in Fig. \ref{fig9num}, where it is clear that this phenomenological model matches with the instabilities and the resonances in a cavity with given geometric and optical parameters. For example, for $V'<1.027$ there is one instability and for $1.027\leq V'< 2.054$ there are two. Similarly, for $V'< 0.513$ there are no resonances and for $0.513\leq V'< 1.54$ there is one resonance. We found a linear behavior given by
\begin{equation}
V_m'\simeq \Delta V' m, \qquad m=0,1/2,1,3/2,2,\dots,
\end{equation}
where $\Delta V'= 1.027$ is the $V'$-size of a single mode. For $m$ integer there are new instabilities, and for $m$ half-integer there are new resonances. Then, the number of instabilities is
\begin{equation}\label{mV}
N_\text{ins}= 1 + \left\lfloor \frac{\omega'_0\sqrt{2 n_0}}{\Delta V'}\tau_\text{c}\sqrt{\delta n_\text{max}}\right\rfloor.
\end{equation}
Comparing Eqs. \eqref{mA} and \eqref{mV}, we can relate the mode area and the $V'$-size as
\begin{equation}
S=\frac{\Delta V'}{\omega'_0\sqrt{2 n_0}}.
\end{equation}

This shows a simpler way to predict the number of instabilities and resonances for a given geometry of the cavity. It is similar to the way that the number of fiber modes is found in the original use of the phenomenological model with $V$. Several $V'_m$ are shown in red dotted lines in Fig. \ref{fig9num}, where they match the instabilities and resonances. From here it is clear that the resonance and instability conditions are related, alternating in integer and half-integer values of $V'_m$.

\subsection{Propagation time}
Up to now, we have verified that the cavity modes are square-integrable in the delay $\tau$ and that $\omega'_R$, the real part of $\omega_\ell'$, fulfills the resonance condition; now we want to focus on the information given by its imaginary part $\omega'_I$. According to Eq. \eqref{exps}, this quantity describes the amplification rate of the instability in terms of the propagation time $\zeta$. In a cavity with more than one instability, the one with the highest $\omega'_I$ is amplified faster and quickly dominates over other instabilities.

We can set numerical values and calculate the propagation time by considering that the model stops working when the peak energy of the quantum fluctuation is that of the classical cavity that holds it, this is our figure of merit. In fact, the energy of the quantum fluctuations is taken from the cavity, as in the Hawking process the energy for the photon production is taken from whatever causes the curvature or surface gravity. In the astrophysical case this is the mass of the black hole, and in optical analogs it is the light pulses causing the optical contrast $\delta n$ of the cavity. This means that the model stops working before this point, but this figure of merit gives us the order of magnitude of the corresponding amplification time $\zeta_\text{a}$ and distance $x_\text{a}$.

The quantum fluctuation in Eq. \eqref{econdaplana2} can be rewritten as
\begin{align}
\phi(\zeta,\tau)=\text{e}^{-i(\omega'_R+i\omega'_I)\zeta}\phi(\tau)
\end{align}
such that
\begin{align}
|\phi(\zeta,\tau)|^2=\text{e}^{2\omega'_I\zeta}|\phi(\tau)|^2
\end{align}
Due to energy conservation in $\zeta$, the total energy density in the system $\rho_\text{S}$ must be the same at any time $\zeta$. Then
\begin{equation}\label{aa}
\rho_{\text{S}}(0)=\rho_\ell(\zeta)+\rho_\text{c}(\zeta),
\end{equation}
where $\rho_\ell$ is the part of the instability
\begin{equation}\label{bb}
\rho_\ell(\zeta)=\hbar \omega_\ell|\phi_{\text{max}}|^2\text{e}^{2\omega_I'\zeta},
\end{equation}
and we take $\omega_\ell=\text{Re}(\omega_\text{2ul})$. The term $\rho_\text{c}$ is the energetic part of the cavity formed by light pulses
\begin{equation}\label{cc}
\rho_\text{c}(\zeta) =\rho_\text{c}(0)\text{e}^{-2\omega_{I}'\zeta},
\end{equation}
where $\rho_\text{c}(0)$ is the initial energy density of the cavity that gives energy to the fluctuation. To find it, recall from Eq. \eqref{n2} that $\delta n_\text{max}$ is related to the maximum intensity of the pulse by $I=\delta n_\text{max}/n_2$. Then
\begin{equation}\label{ee}
\rho_c(\zeta)=\frac{\delta n_{\text{max}}}{n_2}A\text{e}^{-2\omega_{I}'\zeta},
\end{equation}
where $n_2$ is the nonlinear index and $A$ is the effective mode area of the fiber. Comparing the amplification in $\zeta$ of the lasing mode $\rho_\ell$ in Eq. \eqref{bb} and of the cavity $\rho_c$ in Eq. \eqref{ee}, we find that
\begin{equation}
\zeta_\text{a}=\frac{1}{2\omega'_{I}}\ln\left(\frac{\delta n_\text{max}A}{\hbar\omega_\ell|\phi_\text{max}|^2n_2}\right),
\end{equation}
is the lifetime of the instabilities. The distance traveled by the mode inside the optical fiber in that time is $x_\text{a}=u\zeta_\text{a}$. A photonic-crystal fiber commonly used in analog Hawking radiation experiments has nonlinear index $n_2=2.2\times10^{-20}\text{m}^2/\text{W}$ \cite{Agrawal2013}, effective area $A=7.84\times10^{-14} \text{m}^2$, and group velocity $u=2.014\times10^{8}\text{m}/\text{s}\simeq 0.672c$ \cite{Bermudez2016jp}. The amplification distances $x_\text{a}$ are reported in the last column in Table \ref{table1}, where the order of magnitude is of meters. This means that the self-amplified Hawking radiation should obtain high levels of energy, comparable to classical ones, while it propagates in a fiber shorter than a meter.

\section{Conclusions}\label{sec6conclu}
In this work, we studied the propagation of quantum fields in optical fibers and obtained the dispersion relation in the comoving frame for a quantum fluctuation. A light pulse propagating inside a fiber creates analogs of the event horizon. Under certain conditions, a fluctuation is generated spontaneously in an amplification process similar to that of the Hawking radiation in an astrophysical black-hole: This is the optical analog of the Hawking effect.

The resulting fluctuation can be further self-amplified if it is generated in a configuration known as optical black-hole laser, where two pulses trap and amplify the radiation using the energy of the light that forms the cavity. This configuration is usually described by plane-wave modes with real frequencies. 

Here we used the theory of instabilities and allowed all frequencies to be complex. We restricted ourselves to solutions with positive real part of the comoving frequency to describe an amplification process and found the complex laboratory frequencies that allow square-integrability in delay time. With this, we obtained the instabilities as the global cavity modes that are normalizable. The instabilities are trapped inside the cavity and their probability distributions behave like eigenstates in quantum mechanics, for example, the increasing number of maxima in the unstable modes shown in Fig. \ref{fig7prob} and the fact that the confinement increases for a larger cavity. 

We described the evolution of self-amplifying radiation inside the cavity and found some properties that are usually hidden in more complicated models \cite{GaonaReyes2017,Faccio2012}. For example, the fact that analog Hawking radiation can tunnel out of the cavity, similar to the usual behavior of Hawking radiation.

We checked our results for the instabilities by comparing them with the resonance condition in the plane-wave model. The plane-wave model follows the same behavior as the instabilities, although it is more restrictive. Then, we used the instability condition to study the birth-process of an instability and confirmed our previous numerical results. In particular, there is at least one instability for any cavity.

Furthermore, we used a phenomenological model inspired by fiber optics to predict the number of resonances and instabilities. This model matched almost perfectly with the resonance and instability conditions and can be obtained simply with the size and optical contrast of the cavity and the frequency of the initial Hawking radiation. Lastly, we obtained the order of magnitude for the lifetime of the instabilities with a simple model.

A deeper understanding of the parameter region where the number of instabilities and resonances do not match is needed. We believe it is possible to improve the resonance condition to include a dependence on the height of the cavity $\delta n$ as in other models \cite{michel2013saturation,GaonaReyes2017}. Further study on the imaginary part of the instabilities would help us design a cavity with the highest amplification rate of Hawking radiation. The peaks in $\omega'_I$ in Fig. \ref{fig10reim} describe the cavities with the largest amplification. One thing is certain: The theory of instabilities is a powerful tool that can be used in the study of resonant Hawking radiation.

\section*{Acknowledgments}
The authors dedicate this work to the memory of Prof. Renaud Parentani. The authors would like to acknowledge the valuable comments from the anonymous reviewers. JRE acknowledges the financial support of Conacyt (Mexico) through scholarship 637736. The authors acknowledge the financial support of Secretar\'ia de Educaci\'on P\'ublica (Mexico) and Centro de Investigación y de Estudios Avanzados, project 60-2018.

\bibliography{Refs63}
\end{document}